\documentclass[twocolumn]{aastex631}

\usepackage{amsmath,amssymb}
\usepackage{graphicx}
\usepackage{xspace}
\usepackage{orcidlink}

\newcommand{\Msun}{M_{\odot}}
\newcommand{\Rg}{R_{\mathrm{g}}}
\newcommand{\mBH}{m_{\mathrm{BH}}}
\newcommand{\astar}{a_{*}}
\newcommand{\dL}{d_{\mathrm{L}}}
\newcommand{\Rsn}{\mathcal{R}_{\mathrm{Ib}}}
\newcommand{\Pcoinc}{P_{\mathrm{coinc}}}
\newcommand{\FAPcond}{\mathrm{FAP}_{\mathrm{cond}}}
\newcommand{\FAPuncond}{\mathrm{FAP}_{\mathrm{uncond}}}
\newcommand{\pGW}{p_{\mathrm{GW}}}
\newcommand{\pEM}{p_{\mathrm{EM}}}
\newcommand{\kmsMpc}{\mathrm{km\,s^{-1}\,Mpc^{-1}}}

\newcommand{\CO}{\texttt{CO}\xspace}
\newcommand{\COone}{\texttt{CO$1$}\xspace}
\newcommand{\COtwo}{\texttt{CO$2$}\xspace}

\shorttitle{A collapsar-disk origin for GW190814}
\begin{document}

\title{A Collapsar-disk Origin for GW190814}

\newcommand{\columbia}{Department of Physics and Columbia Astrophysics Laboratory, Columbia University, New York, NY 10027, USA}

\author{Vishal Baibhav$\,$\orcidlink{0000-0002-2536-7752}}
\altaffiliation{\href{mailto:vb2630@columbia.edu}{vb2630@columbia.edu}, NASA Einstein Fellow}
\affiliation{\columbia}

\author{Brian D.~Metzger\orcidlink{0000-0002-4670-7509}}
\affiliation{\columbia}
\affiliation{Center for Computational Astrophysics, Flatiron Institute, 162 5th Avenue, New York, NY 10010, USA}

\author{Lam Hui\orcidlink{0000-0001-7003-4132}}
\affiliation{\columbia}

\begin{abstract}

GW190814 was a remarkable gravitational-wave (GW) event: a merger between a $23 M_\odot$ black hole (BH) and a $2.6\,M_\odot$ compact object, with an extreme mass ratio that is difficult to reproduce through standard isolated-binary or dynamical formation channels. Recent work has shown that neutrino-cooled collapsar disks can become gravitationally unstable and fragment, producing neutron stars (NSs) or low-mass BHs in orbit around the newly formed central BH. These fragments may subsequently interact, scatter, merge with one another, or inspiral into the central remnant. We {explore the possibility} that GW190814 originated from such a collapsar-disk fragment merging with the central BH. A key prediction of this scenario is a temporal association with a stripped-envelope supernova preceding the GW event, and we identify the Type Ib supernova candidate SN2019npv, which occurred inside the GW190814 credible volume approximately 60 days before coalescence, as a possible electromagnetic precursor. Although this delay is too long for a conventional kilonova counterpart, we show that three-body interactions among disk fragments can excite some compact objects to wide orbits and produce merger delays of weeks to months. While GW190814 itself was not expected to produce detectable tidal-disruption-powered emission, future delayed mergers in this channel could generate luminous transients through either reprocessed kilonova heating or shocks driven as merger ejecta collide with the preceding supernova ejecta. Finally, treating SN2019npv as the electromagnetic counterpart makes GW190814 a bright standard siren and yields $H_{0} = 71.1^{+9.3}_{-7.5}\,\kmsMpc$.

\end{abstract}

\section{Introduction}
\label{sec:intro}

\begin{figure*}[!htbp]
\centering
\includegraphics[width=\linewidth]{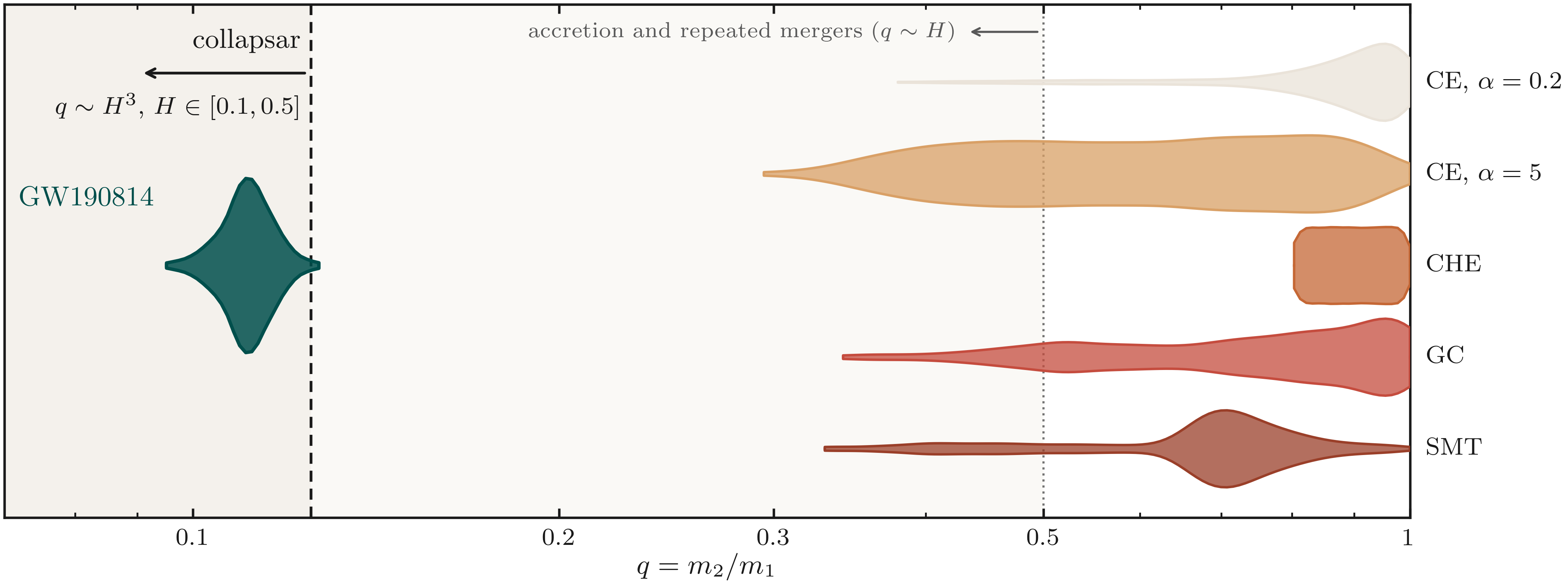}
\caption{Mass ratio of GW190814 compared with predictions from conventional
binary--black-hole formation channels. The teal violin is the GW190814
posterior on $q = m_{2}/m_{1}$; the right-hand violins show theoretical
distributions for common-envelope evolution at efficiencies $\alpha = 0.2$
and $\alpha = 5.0$ (CE), chemically homogeneous evolution (CHE),
globular-cluster dynamics (GC), and stable mass transfer (SMT). The shaded
band at the left marks the collapsar-disk window $q \in [0.001,\,0.125]$,
corresponding to $q \sim H^{3}$ for disk scale heights
$H \in [0.1,\,0.5]$; the dotted line at $q \sim H$ indicates the broader
range available to mergers involving BHs or hierarchical-merger
remnants. GW190814 lies inside the collapsar window and is separated
from every conventional channel, all of which prefer $q \gtrsim 0.5$.
Violins show 99\% credible intervals; channel distributions are adapted from
\citet{Zevin:2020gma}.}
\label{fig:pq}
\end{figure*}

\begin{figure*}[t]
\centering
\includegraphics[width=\linewidth]{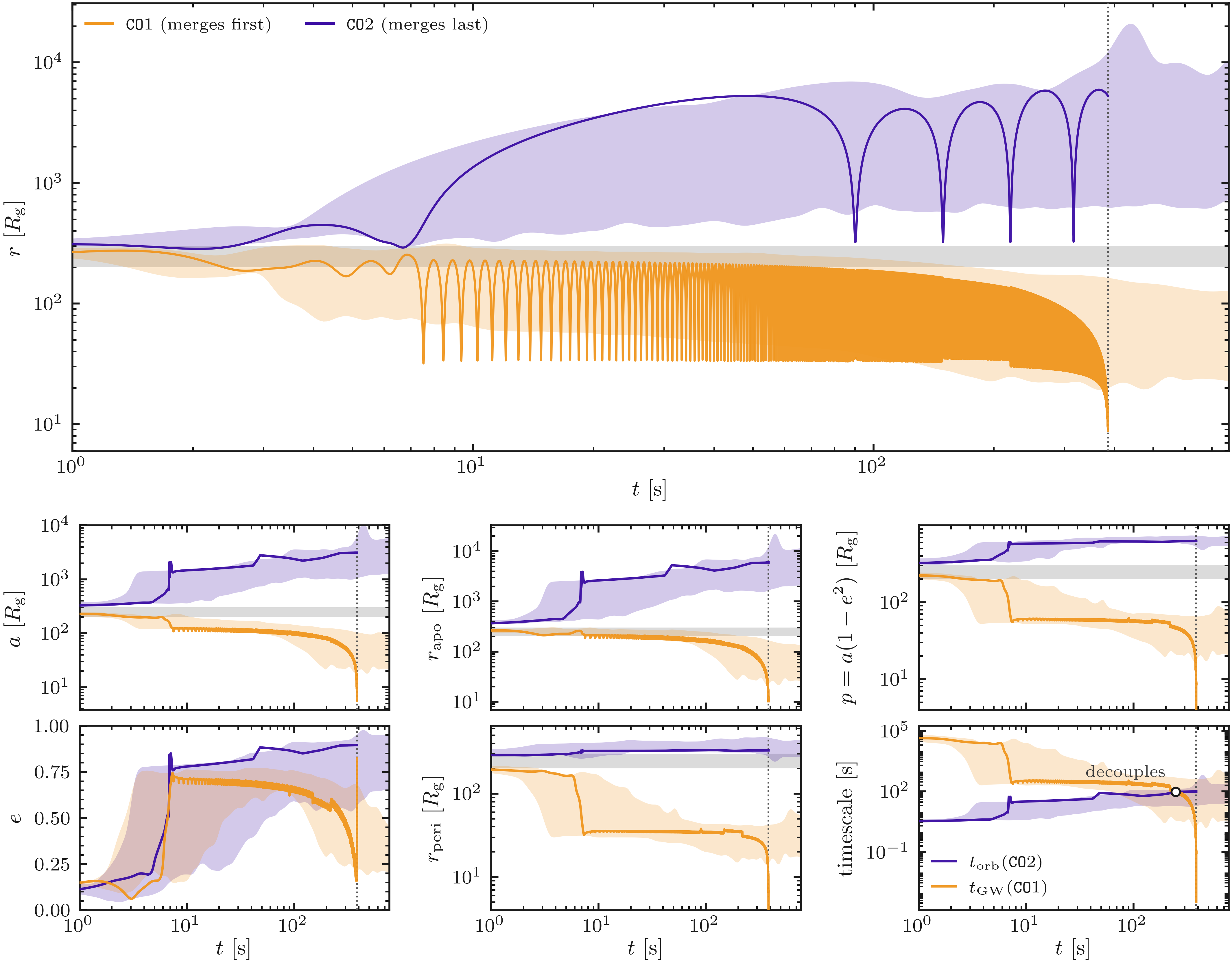}
\caption{{A representative realization of the surviving-companion family
(solid lines), with $5$th--$95$th percentile bands over the ensemble of such
realizations. \COone\ (orange) plunges into the BH first; \COtwo\ (purple)
survives and is the GW190814 progenitor. \emph{Top:} distance from the BH.
\emph{Bottom panels:} orbit-averaged semimajor axis, eccentricity, apocenter,
pericenter, semilatus rectum $p=a(1-e^{2})$, and the two competing
timescales, the orbital period of \COtwo\ and the GW inspiral time of
\COone. The gray band marks the birth annulus $[200,300]\,\Rg$. Within a few
orbits, the conjunctions raise the semimajor axis of \COtwo\ to $a\gtrsim10^{3}\,\Rg$ and
its eccentricity to $e\approx0.8$--$0.9$, while its pericenter remains at the birth radius and its
semilatus rectum remains near $9\bar a/4$ (Equation~\ref{eq:ridge}). \COone\ reaches
a semimajor axis near $\bar a/2$ semilatus rectum $\bar a/4$. The exchange terminates when
$t_{\rm GW}(\COone)$ drops below $t_{\rm orb}(\COtwo)$ (Equation~\ref{eq:freeze}) and \COone\ plunges (dotted line).}
}
\label{fig:time_evol}
\end{figure*}

GW190814 is among the more unusual events in the LIGO--Virgo catalog
\citep{LIGOScientific:2020zkf}. The primary mass is $23.2^{+1.1}_{-1.0}\,M_\odot$ and the secondary is $2.59^{+0.08}_{-0.09}\,M_\odot$, giving a mass ratio $q \equiv m_2/m_1 \approx 0.11$, the lowest yet observed in a confidently detected merger, with the secondary inside the lower mass gap, detected at $\dL = 241^{+41}_{-45}\,\mathrm{Mpc}$. The local rate is
$R = 7^{+16}_{-6}\,\mathrm{Gpc^{-3}\,yr^{-1}}$, and the source sits
between the populations of stellar-mass binary black holes (BHs) and
neutron star (NS) coalescences.

{Conventional channels struggle to reproduce} GW190814's joint combination of mass ratio,
secondary mass, and rate; each fails through its own mass-equalizing physics
(Figure~\ref{fig:pq}). In isolated binaries, stable mass transfer carries
material from the heavier to the lighter star and drives $q$ toward unity
\citep{Broekgaarden:2022nst}, while the common-envelope channel additionally
requires the first Roche-lobe overflow to be stable --- a condition that
$q \approx 0.1$ progenitors generically fail, merging prematurely
\citep{Marchant:2021hiv}; both predict $q \gtrsim 0.5$ and underpredict the
GW190814-like rate by 1--2 orders of magnitude
\citep{Zevin:2020gma}. {Isolated binaries produce GW190814-like systems only through uncommon or fine-tuned physics, such as explodability
fluctuations of massive cores or favorably oriented natal kicks
\citep{Antoniadis:2021dhe,MaganaHernandez:2025sra}.} Chemically
homogeneous evolution also forbids small $q$, since tidal locking and
rotational mixing require comparable masses \citep{Mandel:2015qlu,Marchant:2016wow}. Dynamical assembly in
globular and nuclear star clusters equalizes masses very strongly through
mass segregation and exchange, with NS--BH rates 3--4
orders of magnitude too low \citep{Ye:2019xvf}. Active galactic nucleus (AGN) gas disks could produce
unequal pairings~\citep{McKernan:2020lgr,Tagawa:2020qll} but at higher primary masses~\citep{Yang:2019okq} and uncertain rates. In a
compact hierarchical triple, the secondary could itself be the remnant
of a binary NS (BNS) merger that later merges with a BH tertiary
\citep{Lu:2020gfh}, though this requires the outer orbit to be within
a few astronomical units at the time of the NS coalescence. Population analyses reach the same
conclusion from the data side: including GW190814 destabilizes parametric
mass models of the binary-BH population, and its secondary is assigned a
very low probability of being drawn from the bulk population
\citep{Essick:2021vlx}. Figure~\ref{fig:pq} makes the tension explicit:
GW190814's mass-ratio posterior lies well outside every conventional channel.

A rotating massive star at collapse, however, can produce such binaries. \citet{Metzger:2024ujc} and \citet{Piro&Pfahl07}
have argued that the neutrino-cooled accretion disk surrounding the
nascent BH is gravitationally unstable and cools rapidly
enough to fragment, producing compact objects directly within the
orbit of the central BH; \citet{Chen:2025uwd} confirmed this
picture using 3D shearing-box simulations (see also \citealt{Lerner+26}). Therefore, a collapsar
could produce a $\sim 20\,\Msun$ BH, with disk fragments collapsing to
NSs or low-mass BHs depending on local conditions.
A low-mass compact object in orbit about the central remnant is, by
construction, a small-mass-ratio binary; something that cannot be easily explained by traditional formation channels. We argue that GW190814 could be the
inspiral of one such $\sim 2\,\Msun$ fragment into the central BH. What fragment mass a collapsar disk delivers to the central BH
depends on how far a clump grows before it inspirals, and this is controlled
by the disk aspect ratio $H$. Two scales bracket the possibilities: the Jeans mass of a single unstable clump, $M_{\rm J} \sim H^{3} m_{1}$; and the total
Toomre-unstable disk mass delivered by accretion or repeated mergers, $M_{Q} \sim
H\, m_{1}$. The secondary mass ratio therefore lies between $q \sim H^{3}$
(a single fragment) and $q \sim H$ (resulting from growth via accretion or repeated mergers),
\begin{equation}
H \;\sim\; q^{1/3} \;\text{--}\; q^{1},
\label{eq:HoR}
\end{equation}
so GW190814's $q = 0.11$ implies a thick disk with $H \simeq 0.11$--$0.48$,
as expected for a neutrino-cooled accretion flow \citep{Chen:2006rra}.

{
Fragmentation requires the gas to cool faster than it orbits, a
condition met only outside a fragmentation radius $\astar$ of a few
hundred $\Rg \equiv Gm_1/c^{2}$ \citep{Metzger:2024ujc,Chen:2025uwd}. Inside
$\astar$, the disk remains intact and continues to accrete
onto the BH. Outside it, gravitational instability converts most of the
gas in the unstable annulus into bound fragments. The fragments
therefore form in a region whose density has decreased by orders of
magnitude, and orbit a central mass that consists of the BH and the inner disk. The Jeans mass above
is obtained by integrating the Jeans criterion through a column of the
disk
\citep{Baibhav:2026iny}. However, these fragments can grow by accretion and by merging
with one another~\citep{Lerner+26}. The supply of several solar masses with sufficient
angular momentum to circularize at $\astar$ after the formation of a
$\sim 15$--$20\,\Msun$ BH depends on angular-momentum transport during
the final stages of stellar evolution, which remains poorly
understood. The conditions conducive to fragmentation might
differ from those conducive to a powerful gamma-ray-burst jet. Thus,
``collapsar'' here means the collapse of a rapidly rotating star and
not necessarily a burst progenitor.}

An implication of this picture is an electromagnetic precursor. A collapsar is, by definition, the death of a rotating massive star, and so it should be accompanied by an optical transient. A collapsar progenitor could shed some or most of its hydrogen and possibly
helium envelope by the time it collapses. The channel therefore predicts a
stripped-envelope supernova ---  Ib, Ic, or Ic-BL, sometimes
accompanied by a long gamma-ray burst (LGRB)---for every merger it
produces; the gravitational-wave (GW) event must follow that supernova.

A Type Ib supernova, SN2019npv, was flagged as a candidate counterpart
inside the $90\%$ credible 3D volume of the GW190814 sky map, in
WISEA J005332.35$-$234955.8 at $z = 0.055$ \citep{Ackley:2020qkz}; a spectrum
obtained $\sim 12\,$days after the trigger matched a Type Ib supernova
$\sim 50\,$days past peak brightness \citep{Andreoni:2019qgh}. Converting this to
an explosion epoch requires only the time a Type Ib takes to rise from
explosion to optical peak, measured to be $\sim 22\,$days in well-sampled
stripped-envelope samples \citep{Drout:2010ww,Bianco:2014mna,Taddia:2017rvw}; the
supernova therefore exploded $\sim 60\,$days before the GW
merger. Several follow-up campaigns ruled out SN2019npv as a GW190814
counterpart \citep{Vieira:2020lze,Watson:2020iqj,Andreoni:2019qgh,GravityCollective:2021kyg}.
Those conclusions, however, assumed that an electromagnetic counterpart
(presumed to be a kilonova) should arrive at, or near, the moment of
coalescence. A $\sim 60\,$day separation is causally impossible for a
kilonova counterpart but is broadly consistent with the merger delay expected if the supernova is the collapsar that later delivers a fragment to the central
BH (Section~\ref{sec:nbody}).

{\textit{Scope of this Letter.} No global simulation yet follows a
collapsar disk self-consistently from a collapsing stellar profile
through fragmentation; the shearing-box calculations establish the
local mechanism, given a marginally unstable, rapidly cooling patch~\citep{Chen:2025uwd},
but not that a given progenitor produces such a patch after forming a
$\sim 15$--$20\,\Msun$ BH. The production of such a patch by a specific
progenitor remains unresolved and is outside the scope of this work.
We therefore adopt the existence of a fragmenting
outer disk as the working hypothesis, state the conditions it imposes
on the progenitor and on the mass and spin of the central BH
(Section~\ref{sec:nbody}, Appendix~\ref{sec:rate}), and derive its
observational consequences. Accordingly, the analysis is conditional
on this hypothesis and yields a specific, testable sequence of
observables. We do not perform hydrodynamic or GRMHD simulations: the
few-body integrations are carried out in vacuum, and the effects of the
gas on the fragment trajectories and on the spinup and spindown of
the BH are estimated from published results (Section~\ref{sec:nbody}).}

The remainder of this Letter develops the scenario and its
consequences. We first show that scattering among disk fragments
produces the weeks-to-months delay demanded by the $\sim 60\,$day
SN2019npv offset (Section~\ref{sec:nbody}). We then exploit the
candidate host to measure $H_{0}$ (Section~\ref{sec:h0}), and we show that future delayed mergers in this channel can power distinctive ``embedded'' transients when the merger ejecta are reprocessed by, or collide with, the preceding supernova ejecta (Section~\ref{sec:embedded_kn}). Two further
checks are presented in the appendices: that the collapsar rate can plausibly supply such events (Appendix~\ref{sec:rate}), and that the SN2019npv
association has a chance alignment of $\sim2\sigma$
(Appendix~\ref{sec:host}).

\begin{figure*}[!htbp]
\centering
\includegraphics[width=0.9\linewidth]{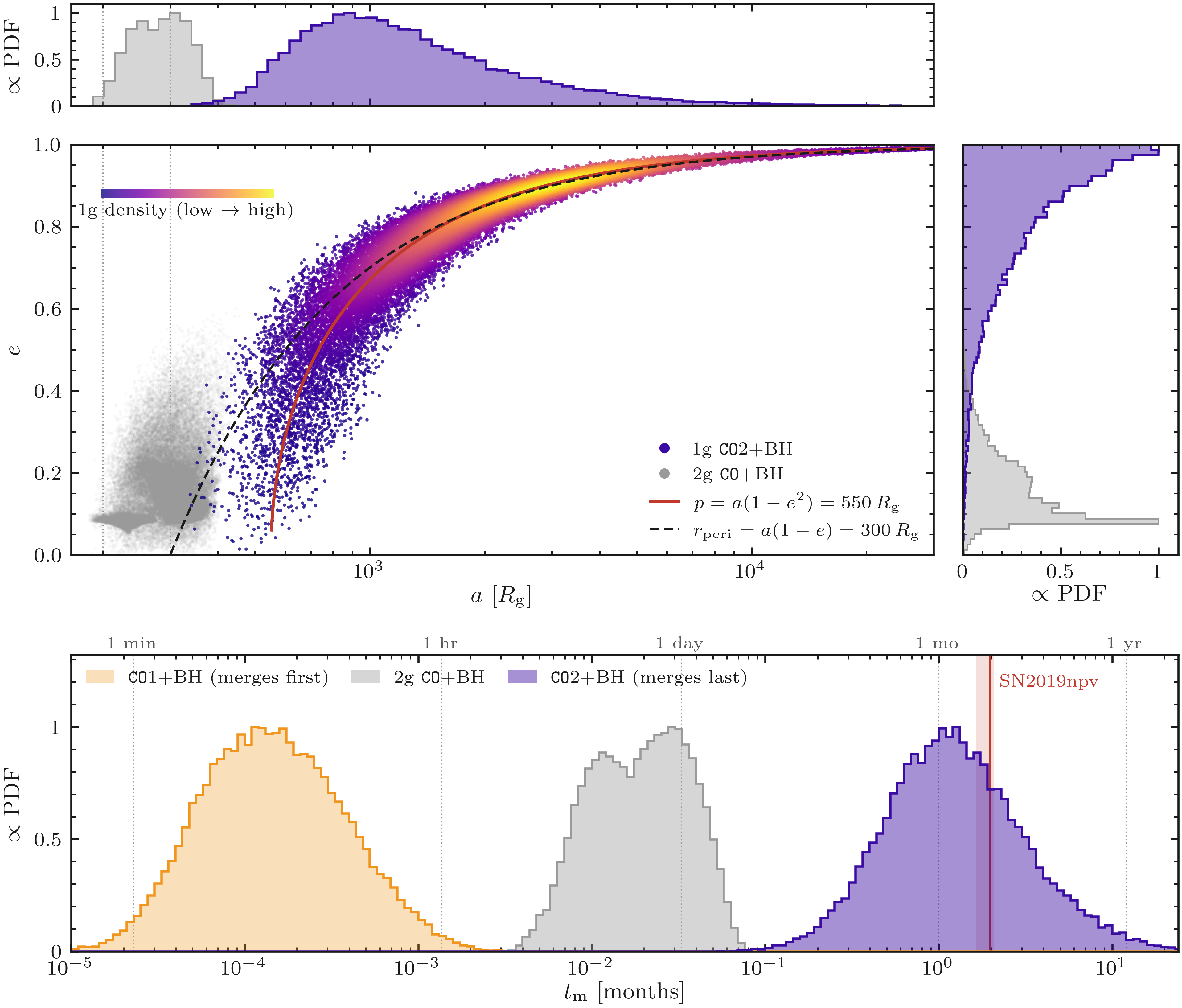}
\caption{\emph{Top panels:} joint distribution of post-first-event
semimajor axis $a$ and eccentricity $e$,
with marginal histograms. The surviving outer companion (1g \COtwo$+$BH; colored
density) traces a high-eccentricity, large-$a$ ridge with $e \to 1$ and
$a \sim 10^{3}$--$10^{4}\,\Rg$, while the second-generation product
(2g \CO$+$BH; gray) clusters at $a \lesssim$ few $\times 10^{2}\,\Rg$ with
modest eccentricity. \emph{Bottom panel:} distribution of merger delays
between the supernova precursor and the GW event. The delay
is multimodal: the inner companion \COone, which merges first, plunges within
minutes to hours; 2g~CO$+$BH products span minutes to days; and the surviving
outer companion (1g~\COtwo$+$BH) peaks at $\sim 30$~days and extends to
roughly a year. The vertical orange line marks the $\sim 60$-day
delay (plausibly $50$--$63$ days) between SN2019npv and GW190814, which falls inside the
delayed-merger peak.}
\label{fig:corner}
\end{figure*}

\begin{figure}[!htbp]
\centering
\includegraphics[width=\linewidth]{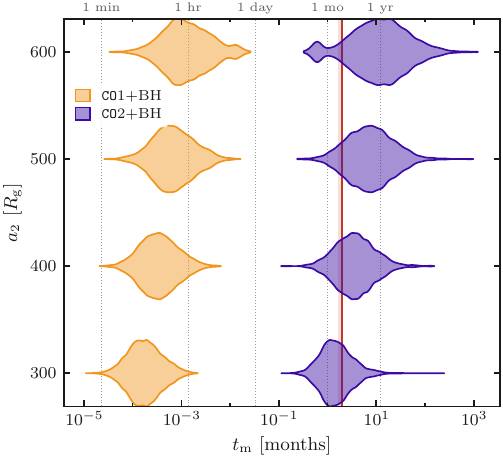}
\caption{{Merger delays for compact objects sampled from a
fragmenting region with inner edge $a_{1}=200\,\Rg$ and outer edge
$a_{2}=300$, $400$, $500$, or $600\,\Rg$. Each row shows the delays of
the prompt \COone$+$BH mergers (orange) and the delayed \COtwo$+$BH
mergers (purple). The red line denotes the $60$\,day offset between SN2019npv and GW190814.}}
\label{fig:robust}
\end{figure}

\begin{figure*}[!htbp]
\centering
\includegraphics[width=0.72\linewidth]{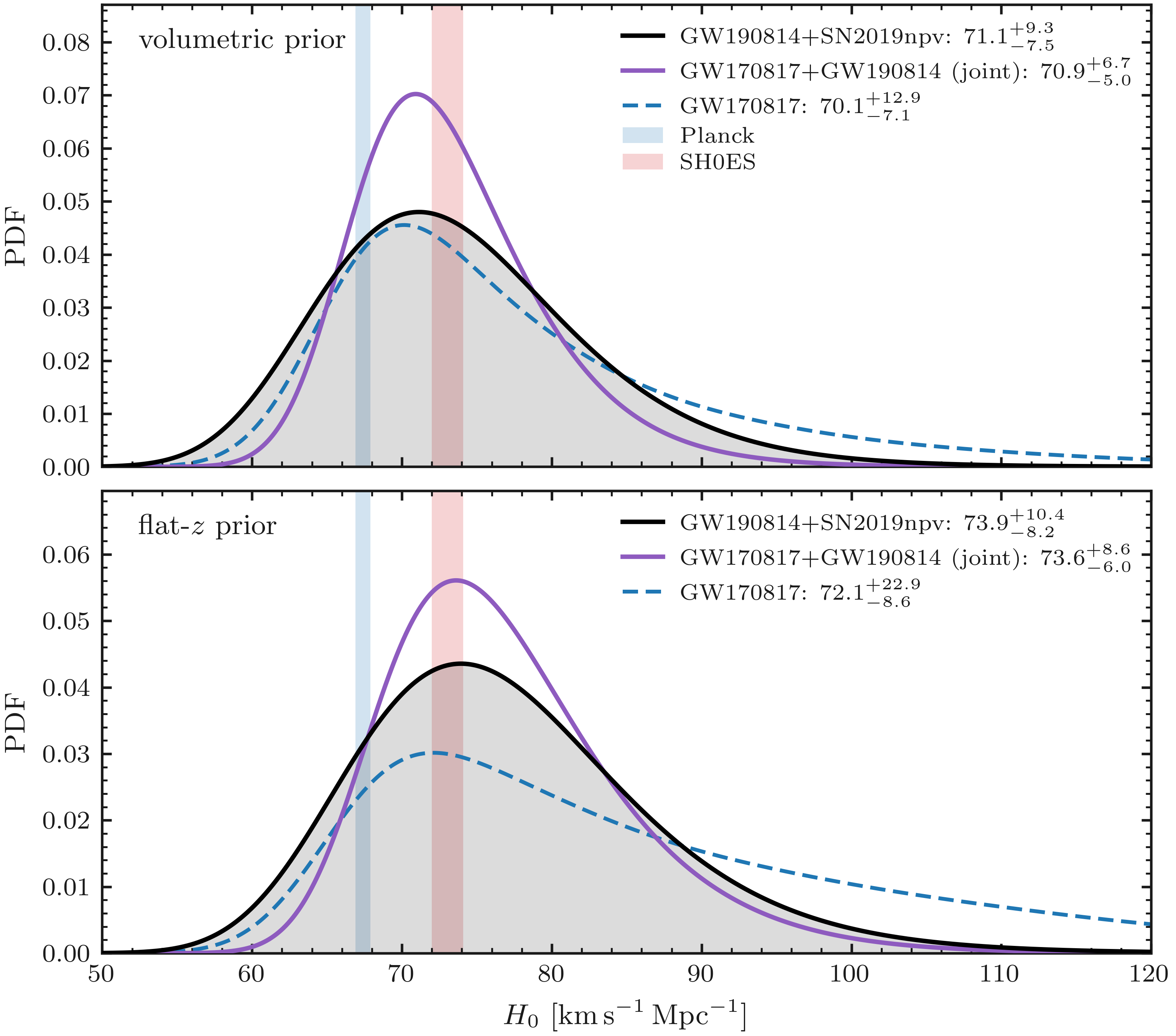}
\caption{Posterior on $H_{0}$ obtained by treating SN2019npv as the
electromagnetic counterpart preceding GW190814, shown for the two
redshift-prior conventions. The top panel shows the volumetric prior, and the bottom panel shows the flat-in-redshift prior. Each panel shows GW190814
(black, shaded) and GW170817 \citep[blue dashed;][]{LIGOScientific:2017adf};
their joint posterior (purple) is the product of the two single-event
likelihoods under a single $H_0$ prior, $\pi(H_0)\propto 1/H_0$, and it assumes
SN2019npv is the true host (see Appendix~\ref{sec:host} for the
chance-coincidence statistics). Labels give the maximum a posteriori value and
minimal $68.3\%$ credible interval; vertical bands mark Planck~2018 (blue)
and SH0ES (red).}
\label{fig:h0}
\end{figure*}

\section{Delayed Mergers from Scattering}
\label{sec:nbody}

A fragmenting collapsar disk will in general produce more than one
clump \citep{Chen:2025uwd}. A system of multiple gravitating bodies orbiting a central mass
on closely spaced, near-coplanar orbits is generically dynamically
unstable. Such a system becomes unstable on a few crossing times, with
outcomes that depend only weakly on initial conditions: mutual
scattering, ejection, or close encounter with the central mass
\citep{1996Sci...274..954R,1996Icar..119..261C,Ford:2007ea,Chatterjee:2007ed,
Juric:2007dx,2009Icar..201..381S,2015ApJ...807..157T,2017A&A...605A..72L,Naoz:2016cjb}.
Multiple collapsar-disk fragments behave similarly. The only
modification, at $\sim 200\,\Rg$ from a $20\,\Msun$ BH, is
that close encounters are mildly relativistic, and GW
emission is nonnegligible{; the role of the residual disk gas is
addressed at the end of this section}.

{We simulate a nonspinning central BH of mass $\mBH = 20.4\,\Msun$ and two $2.6\,\Msun$ compact objects ({\CO}s), on coplanar, initially circular orbits with semimajor axes drawn uniformly in area from $[200,300]\,\Rg$. The calculation includes neither disk mass nor accretion, and the merger of the BH with one fragment produces a $23\,\Msun$ primary. Mass accretion and unequal fragment masses are considered separately at the end of this section. The $2.6\,\Msun$ secondary may form with that mass in a disk of scale height $H\sim0.5$, or attain it through accretion or an earlier merger with another fragment; only its mass at the final inspiral is constrained by GW190814.}
 These radii are typical of those expected to undergo gravitational instability and fragmentation in collapsar disks \citep{Metzger:2024ujc}{, and the width of the annulus, $\sim H a$, is one Jeans length; the fragmenting zone can be wider, which we examine below (Figure~\ref{fig:robust})}. We integrate with AR-CHAIN
\citep{2008AJ....135.2398M} including post-Newtonian corrections through
3.5\,PN. We label the two {\CO}s: \COone and \COtwo.
The full distribution of disk fragments will be richer than two
bodies; two is the smallest experiment that can settle the question of
merger-delay statistics. We find two distinct subpopulations across our $10^{5}$ Monte Carlo experiments:

\noindent\textit{Prompt inner $+$ delayed outer merger.}
In the first family, \COone and \COtwo mutually scatter. The repeated
interactions send \COone toward the central BH; \COtwo is
simultaneously kicked outward, with its semimajor axis growing and
its eccentricity pumped toward unity. In a purely Newtonian setting,
many would-be escapers acquire $e > 1$ and unbind, as in planet--planet
scattering \citep{Ford:2007ea,Chatterjee:2007ed,Juric:2007dx}.
{GW emission modifies the late-time dynamics: \COone\ decouples from
\COtwo\ and plunges before sufficient energy is transferred to unbind
\COtwo, which therefore remains on a wide, eccentric bound orbit
(see below).} In some cases, however,
\COtwo\ attains $e > 1$ and is
ejected from the system. After repeated encounters, \COone inspirals and merges with the central
BH within minutes to hours. Such a prompt merger was not detected in GW190814 (see Section~\ref{sec:discussion}). Because \COone\ is present only briefly before plunging, it cannot fully eject \COtwo, so the outer companion survives. From the \COone$+$BH merger onward, the
evolution is set entirely by GW-driven inspiral, and \COtwo$+$BH is the
GW190814 progenitor. Figure~\ref{fig:time_evol} shows the resulting
divergent evolution of the two semimajor axes: \COone contracts toward the BH while \COtwo climbs to $\gtrsim 10^{3}\,\Rg${; the figure also identifies the time at which the GW inspiral time of \COone\ becomes shorter than the orbital period of \COtwo\ and the two decouple}. With \COone\ gone and nothing left to pump its eccentricity, \COtwo\ circularizes via GW emission \citep{Peters:1964zz} during the month-long inspiral, entering the LVK band essentially circular. The prompt \COone$+$BH merger, by contrast, occurs at high eccentricity \emph{within} the band and would be a distinctive signature \citep{Wu:2026hth}.

\noindent\textit{Fragment merger $+$ 2g plunge.} In the second family, \COone and \COtwo merge with each other first. These
mergers are prompt, $\mathcal{O}(\mathrm{s})$ after the encounter; the
signal is weak owing to the short duration and low component masses,
and would not be detectable at the GW190814 distance. The merger product is a second-generation (2g) compact object on a bound orbit about the central BH. The 2g object subsequently plunges into the
central BH over minutes to days.

The two subpopulations occupy well-separated regions of the
post-first-event $(a,e)$ plane (Figure~\ref{fig:corner}). The
\COtwo$+$BH systems lie at $a \sim 10^{3}$--$10^{4}\,\Rg$
with $e$ approaching unity; the 2g \CO$+$BH products  sit
at a few hundred $\Rg$ with modest eccentricities. The merger-delay
distribution is correspondingly bimodal (Figure~\ref{fig:corner}):
minutes-to-days for 2g \CO + BH mergers, days-to-year for \COtwo + BH mergers with a
peak at $\sim 30\,$days.

The observed $\sim 60\,$day separation between SN2019npv and GW190814
is consistent with merger of BH with outer companion.

{In the experiment above, both fragments are born within one
annulus of width $\sim H a$, one Jeans length. The fragmenting zone
can be wider, depending on how far out the disk cools efficiently, and
the two fragments need not be born at the same radius.} {We
therefore repeat the experiment with both semimajor axes drawn
uniformly in area from wider bands, $[a_{1},a_{2}]$ with
$a_{1}=200\,\Rg$ and $a_{2}=300$, $400$, $500$, and $600\,\Rg$
(Figure~\ref{fig:robust}). The prompt \COone$+$BH mergers occur within
minutes to hours in every band. About one-third of the realizations
end in a delayed \COtwo$+$BH merger, $40\%$--$50\%$ in a
\COone$+$\COtwo\ collision, and the remainder in the ejection of
\COtwo. The median delay of the \COtwo$+$BH merger increases with the
bandwidth, as expected from the dependence of the delay on the birth
radius (Equation~\ref{eq:tdelay}). The $60$\,day offset of SN2019npv
lies within the central $68\%$ of the distribution for
$a_{2}\leq400\,\Rg$ and in its lower part for wider bands.}

A fragmenting collapsar disk will
generally produce more than two clumps; our two-body experiment is
thus the minimal setting in which the merger-delay distribution can
be characterized. We find that the delayed mergers carry over to a
larger number of particles with much richer dynamics and GW
signatures, which we will present in future work.

\subsection{Energy and Angular-momentum Exchange}

{The retention of \COtwo\ and the location of the surviving
companions along the locus $a(1-e^{2})\approx{\rm const}$ in
Figure~\ref{fig:corner} both follow from the exchange of energy and
angular momentum at conjunctions. At a conjunction, the two
fragments receive equal and opposite {momentum impulses, so that
their velocity changes,
$\delta\mathbf{v}=\delta v_{r}\,\hat{\mathbf r}+\delta v_{t}\,\hat{\boldsymbol\phi}$,
are inversely proportional to their masses}.
Only the tangential component changes the specific angular momentum,
$\Delta\ell = r\,\delta v_{t}$, while for nearly circular orbits, or for encounters near apocenter or pericenter where $v_{r}\approx0$,
$\Delta\varepsilon \simeq v_{t}\,\delta v_{t}$; energy and angular
momentum are therefore exchanged in a fixed ratio,
$\Delta\varepsilon/\Delta\ell \simeq \Omega$, the local orbital
frequency. Conjunctions recur on the synodic time,
\begin{equation}
t_{\rm syn}=\frac{2\pi}{|\Omega_{1}-\Omega_{2}|}
\simeq\frac{4\pi}{3\bar\Omega}\,\frac{\bar a}{\Delta a}
\approx 4\,{\rm s}
\label{eq:tsyn}
\end{equation}
for $a_{1}=200$ and $a_{2}=300\,\Rg$ around a $20\,\Msun$ BH,
comparable to the orbital period. The synodic time sets the rate of the
exchange, and the first strong encounter occurs within a few synodic times
(Figure~\ref{fig:time_evol}). {Let $\mu\equiv m_{\COone}/m_{\COtwo}$
be the mass ratio of the two fragments, and} let $x$ be the fraction of its initial {specific} binding energy,
$\varepsilon_{0}=G\mBH/2\bar a$, that \COtwo\ has received from \COone,
with $\bar a$ being the birth radius and $\ell_{0}$ being the {specific} angular
momentum there. {We write the exchange in terms of specific
quantities (per unit mass), because these determine the orbital
elements. The total energy and angular momentum are conserved in the
exchange, so the energy $x\,\varepsilon_{0}\,m_{\COtwo}$ gained by
\COtwo\ is lost by \COone; per unit mass of \COone\ this is
$x\,\varepsilon_{0}/\mu$. The same holds for the angular momentum.
Along the ratio $\Delta\varepsilon/\Delta\ell=\Omega$ established above, the
angular momentum that accompanies an energy $x\,\varepsilon_{0}$ is
$x\,\varepsilon_{0}/\Omega=(x/2)\,\ell_{0}$, since
$\varepsilon_{0}=\tfrac12\bar a^{2}\Omega^{2}$ and
$\ell_{0}=\bar a^{2}\Omega$ for a circular orbit of radius $\bar a$.
Starting from $a_{1}=a_{2}=\bar a$, successive conjunctions then yield the sequence}
\begin{equation}
\begin{aligned}
\varepsilon_{2}&=-(1-x)\,\varepsilon_{0}, & a_{2}&=\frac{\bar a}{1-x}, & \ell_{2}&=\Bigl(1+\frac{x}{2}\Bigr)\ell_{0},\\
\varepsilon_{1}&={-\Bigl(1+\frac{x}{\mu}\Bigr)\varepsilon_{0}}, & a_{1}&={\frac{\bar a}{1+x/\mu}}, & \ell_{1}&={\Bigl(1-\frac{x}{2\mu}\Bigr)\ell_{0}},
\end{aligned}
\label{eq:rail}
\end{equation}
\COtwo\
gains energy and angular momentum and its orbit widens; \COone\ loses
both and its orbit shrinks. {The corresponding eccentricities are
\begin{equation}
1-e_{2}^{2}=(1-x)\Bigl(1+\frac{x}{2}\Bigr)^{2},\qquad
1-e_{1}^{2}=\Bigl(1+\frac{x}{\mu}\Bigr)\Bigl(1-\frac{x}{2\mu}\Bigr)^{2}.
\label{eq:ecc}
\end{equation}
The exchange is bounded in two ways. At $x=1$,
$1-e_{2}^{2}=0$: \COtwo\ has received all of its binding energy,
$a_{2}\to\infty$ and $e_{2}\to1$, and it is unbound. At $x=2\mu$,
$1-e_{1}^{2}=0$: \COone\ has given away all of its angular momentum,
$\ell_{1}=0$ and $e_{1}\to1$, and it falls radially toward the BH.
Per unit energy transferred, \COone\ loses angular momentum at a rate
proportional to $1/\mu$. If $\mu<1$, \COone\ is lighter than \COtwo\
and exhausts its angular momentum after transferring less energy.
The first bound reached is determined by whether $2\mu$ exceeds unity.

\textit{Comparable masses, $\mu\geq1/2$.} The bound $x=1$ is reached
first. For equal masses, $\mu=1$, as in our integrations,
\COone\ then has} $a_{1}=\bar a/2$, $e_{1}=1/\sqrt{2}$, and a
pericenter of $0.15\,\bar a\approx 35\,\Rg$, after transferring half of
its angular momentum. This is the standard outcome of planet--planet
scattering, characterized by one body at approximately half the initial
separation and near-parabolic ejection of the other
\citep{1996Sci...274..954R,Ford:2007ea}. We reproduce it when the
radiative (2.5PN) terms are omitted: delayed mergers are then almost
entirely absent, and \COtwo\ is either ejected or collides with \COone.

However, gravitational radiation---absent in planet--planet
scattering---modifies this outcome. First, conjunctions become infrequent as $x \to 1$ because the period of \COtwo\ grows as
$(1-x)^{-3/2}$. Second, as \COone\ approaches
$a_1 \approx \bar a/2$ with a $\sim35\,\Rg$ pericenter, it becomes an eccentric
extreme-mass-ratio inspiral whose GW decay time is minutes. The exchange terminates when the decay
time of \COone\ drops below the orbital period of \COtwo, which happens
at
\begin{equation}
1-x \simeq 32\,(67\pi)^{2/3}\left[q\left(\frac{\Rg}{\bar a}\right)^{5/2}\right]^{2/3}
\approx 0.025
\label{eq:freeze}
\end{equation}
for $q=0.1$ and $\bar a=250\,\Rg${, where
$t_{\rm GW}=a/|\dot a|$ is the GW decay time}. \COone\ then
decouples and plunges within minutes to hours. \COtwo\ remains
marginally bound, on an orbit with $a_{2}=\bar a/(1-x)\sim
10^{4}\,\Rg$ and $e_{2}\approx0.97$, with its angular momentum
fixed at $\ell_{2}\simeq\tfrac32\ell_{0}$. Its semilatus rectum is
therefore fixed,
\begin{equation}
p_{2}=a_{2}(1-e_{2}^{2})\simeq \frac{9}{4}\,\bar a\approx 560\,\Rg,
\label{eq:ridge}
\end{equation}
which corresponds to the ridge in Figure~\ref{fig:corner}: scattering
distributes the energy of \COtwo\ over an order of magnitude while its
angular momentum remains nearly constant. As shown in Figure~\ref{fig:time_evol}, the semilatus rectum of \COtwo\ increases by a factor of about $9/4$ while its pericenter remains nearly constant; conversely, the semilatus rectum of \COone\ decreases while its apocenter remains nearly constant. Finally, the final orbit of \COtwo,
\begin{equation}
\begin{aligned}
a_{2}&=\frac{(\bar a/\Rg)^{8/3} \,\Rg}{32\,(67\pi)^{2/3}\,q^{2/3}}\simeq10^{4}\,\Rg\left(\frac{\bar a}{250\,\Rg}\right)^{8/3}\left(\frac{q}{0.1}\right)^{-2/3},\\
1-e_{2}^{2}&=\frac{72\,(67\pi)^{2/3}\,q^{2/3}}{(\bar a/\Rg)^{5/3}}
\simeq0.05\left(\frac{q}{0.1}\right)^{2/3}\left(\frac{\bar a}{250\,\Rg}\right)^{-5/3},
\end{aligned}
\label{eq:finalorbit}
\end{equation}
has a GW inspiral time {$t_{\rm m}$. The corresponding birth
radius is
\begin{equation}
\begin{aligned}
\frac{\bar a}{\Rg} &=\left[\frac{348160\sqrt{2}\,(67\pi)^{1/3}\,q^{4/3}}{6561}\,
\frac{c^{3}\,t_{\rm m}}{G\mBH}\right]^{6/29}\\
&\simeq 310
\left(\frac{t_{\rm m}}{60\,{\rm d}}\right)^{6/29}
\left(\frac{q}{0.1}\right)^{8/29}
\left(\frac{\mBH}{20\,\Msun}\right)^{-6/29},
\end{aligned}
\label{eq:tdelay}
\end{equation}}
{The corresponding delay} agrees with the peak of the simulated distribution to within a
factor of a few. The delay depends only on the birth radius and the
mass ratio; the 60\,day offset of SN2019npv corresponds to
$\bar a\approx {310}\,\Rg$. The steep dependence on the birth radius,
$t_{\rm m}\propto\bar a^{29/6}$, together with the statistical
spread in $1-x$, also implies a broad distribution: modest variations
in the birth radii and in the final conjunctions translate into delays
ranging from days to about a year, as seen in Figures~\ref{fig:corner}
and~\ref{fig:robust}.

{\textit{Light plunging fragment, $\mu<1/2$.} If \COone\ is
lighter than half of \COtwo, the bound $x=2\mu<1$ is reached first:
\COone\ exhausts its angular momentum before it has supplied enough
energy to unbind its heavier companion. At this limit,
$a_{1}=\bar a/3$ and $\ell_{1}=0$, so \COone\ plunges in a near-radial
orbit. The exchange terminates with this plunge rather than through
the GW-driven decoupling of Equation~\ref{eq:freeze}. The final orbit
of \COtwo\ is
\begin{equation}
a_{2}=\frac{\bar a}{1-2\mu},\qquad e_{2}^{2}=\mu^{2}(3+2\mu),\qquad
p_{2}=(1+\mu)^{2}\,\bar a .
\label{eq:lightorbit}
\end{equation}
A lighter fragment transfers less energy, so \COtwo\ ends on a more
compact and less eccentric orbit than for $\mu=1$: for
$m_{\COtwo}=2.6\,\Msun$ and
$m_{\COone}=0.5$--$1\,\Msun$ ($\mu\approx0.2$--$0.4$),
$a_{2}\approx(1.6$--$4.3)\,\bar a$ and $e_{2}\approx0.35$--$0.75$.
The Peters GW inspiral time \citep{Peters:1964zz} gives the birth
radius required for a merger time $t_{\rm m}$,
\begin{equation}
\frac{\bar a}{\Rg}=\left[\frac{256}{5}\,
\frac{q\,(1-2\mu)^{1/2}F(e_{2})}{(1+\mu)^{7}}\,
\frac{c^{3}\,t_{\rm m}}{G\mBH}\right]^{1/4},
\label{eq:tdelaylight}
\end{equation}
with $q=m_{\COtwo}/\mBH$ and
$F(e)=1+\tfrac{73}{24}e^{2}+\tfrac{37}{96}e^{4}$. For $q=0.1$ and
$\mBH=23\,\Msun$, the 60\,day offset of SN2019npv requires
$\bar a\approx520\,\Rg$ for $m_{\COone}=0.5\,\Msun$ and
$\bar a\approx420\,\Rg$ for $m_{\COone}=1\,\Msun$, compared with
$\bar a\approx300\,\Rg$ for $\mu=1$ at the same $\mBH$
(Equation~\ref{eq:tdelay}).

This argument does not determine whether the lighter fragment more
often plunges or is ejected. In star clusters, two-body relaxation drives the system
toward energy equipartition: lighter bodies acquire larger velocities
and move outward, while heavier bodies lose energy and sink toward the
center. If fragment interactions have the same mass
dependence, the lighter fragment may more often gain energy and be
ejected, in which case the
configuration with a light plunging fragment is the less frequent
outcome. If so, the fraction of systems in which a light fragment
plunges and the heavier one is retained decreases with decreasing
$\mu$.}

This derivation provides a simplified approximation to the simulated dynamics.
It assumes that the impulses at conjunction are purely tangential. In
practice, the radial components are also important; radial impulses do not change the angular momentum, but
they do change the energy
and increase the eccentricity at fixed $p$. The derivation also neglects
GW emission during the scattering phase, which removes energy and
angular momentum from \COone\ before it decouples, and it assumes that
both fragments start at the same radius, $a_{1}=a_{2}=\bar a$, whereas
in the simulations they are drawn from $[200,300]\,\Rg$. The reflex
motion of the BH, the mutual potential energy of the fragments at
conjunction, and relativistic periapsis precession are likewise
omitted. {The derivation further treats $x$ as a continuous
variable, whereas energy is transferred in discrete increments at
successive conjunctions. The exchange therefore stops at the last
conjunction before the limit is reached, and the value of $x$ implied
by Equation~\ref{eq:freeze} is an upper limit; in the integrations $x$
remains farther from unity. The semilatus rectum,
$p_{2}=(1+x/2)^{2}\,\bar a$, depends only weakly on $x$ near $x=1$, and
Equation~\ref{eq:ridge} accordingly reproduces the ridge in
Figure~\ref{fig:corner}; the semimajor axis, $a_{2}=\bar a/(1-x)$,
depends strongly on $1-x$, and the values of $a_{2}$ in
Figure~\ref{fig:corner} are correspondingly smaller than those given by
Equation~\ref{eq:finalorbit}.} Equations~\ref{eq:rail}--{\ref{eq:tdelaylight}} should therefore be
regarded as estimates that identify the relevant scalings; the
distributions shown in Figures~\ref{fig:corner} and~\ref{fig:robust}
are taken from the full integrations.}

\subsection{Disk Gas, Accreted Mass, and the Spin of the Primary}

{The few-body calculation above is carried out in vacuum, with
$m_{1}=\mBH+m_{\COone}$ and $\mBH=20.4\,\Msun$; it includes neither disk
mass nor accretion. When subsequent accretion is included, the mass
bookkeeping becomes
$m_{1}=m_{\rm BH,0}+\Delta m_{\rm acc}+m_{\COone}$. The
fragments are born in the depleted zone outside $\astar$, where the
residual gas is a small fraction of the pre-fragmentation density.
Gas-driven migration and drag there act on timescales
$\gtrsim10^{2}$--$10^{3}\,$s \citep{Baibhav:2026iny}, far longer than
the $\sim4\,$s synodic time on which the first strong encounter occurs
(Equation~\ref{eq:tsyn})
and the few orbits within which \COtwo\ is scattered onto its wide orbit
(Figure~\ref{fig:time_evol}). Thereafter, its pericenter remains at or
beyond its birth radius (Equation~\ref{eq:ridge}), so \COtwo\ does not
enter the intact inner disk; its subsequent evolution is governed by GW
emission, and the month-long delay is unaffected by the gas. The
situation differs for \COone.  It does affect \COone\ once its
pericenter enters inner disk, where the migration timescale is
$\lesssim0.1\,$s \citep{Baibhav:2026iny}, and its trajectory is
strongly modified.  Gas there removes angular momentum from
\COone, accelerates its plunge, and dephases its GW signal relative to a
vacuum inspiral. We do not model this effect, which may have contributed to the
nondetection of the prompt \COone$+$BH merger
(Section~\ref{sec:discussion}); it does not alter the outcome for
\COtwo, which, by then, is on a wide orbit.

The same accretion determines the final mass and spin of the primary. The
mass of the $23\,\Msun$ primary of GW190814 is the sum of the BH mass at
the time of fragmentation, $m_{\rm BH,0}$, the subsequently accreted
inner-disk mass, $\Delta m_{\rm acc}$, and the mass of \COone. The merger with \COone\ increases
the BH spin. For a
quasi-circular prograde plunge, the resulting spin is
$\chi\simeq 2\sqrt{3}\,q/(1+2\sqrt{2}\,q/3)^{2}$ with
$q=m_{\COone}/m_{\rm BH,0}$, which gives $\chi\simeq0.3$--$0.40$. The post-merger spin {for a plunging
fragment with $m_{\COone}=m_{\COtwo}$}
exceeds the measured $\chi_{1}<0.07$ \citep{LIGOScientific:2020zkf}.
Reproducing the measured value therefore requires{, in that case,} subsequent accretion
of the inner disk in a magnetically arrested state, in which Blandford--Znajek
torques \citep{Blandford:1977ds} drive the spin toward an equilibrium
value $\chi_{\rm eq}$
\citep{Narayan:2021qfw,Lowell:2023kyu,Gottlieb:2023cgm,Jacquemin-Ide:2023aax}.
Existing GRMHD simulations show that thick, hot magnetically arrested
disks robustly spin BHs down to low equilibrium spins, but the precise
value and its dependence on disk thickness, thermodynamics,
magnetic-flux saturation, and field geometry remain poorly calibrated
in the regime relevant here. Thin disks reach
$\chi_{\rm eq}\simeq0.3$ and thick adiabatic disks $\simeq0.07$, with
$\chi_{\rm eq}$ falling steeply with thickness
\citep{Lowell:2023kyu,Lowell:2025prd}; neutrino-cooled collapsar disks
at high accretion rates give $\simeq0.13$ \citep{Issa:2025jzq}.
\citet{Lowell:2025prd} fit $\chi_{\rm eq}\simeq0.3-2.7H^{2}$ to
simulations with $H\le0.3$. The dependence is steep: the fit gives
$\chi_{\rm eq}\approx0.06$ at $H=0.3$ and $\approx0.006$ at $H=0.33$,
and it cannot be extrapolated to the $H\approx0.5$ assumed here, which
has not been simulated.

Thicker and hotter flows, for which the fits imply
$\chi_{\rm eq}\ll0.1$, have not yet been simulated. The late-time
collapsar disk is expected to be in this regime: once the accretion
rate falls below the neutrino-cooling threshold, the inner disk becomes
hot and geometrically thick. The scenario therefore requires efficient
cooling and weak magnetization in the fragmenting outer disk, together
with a thick, magnetically arrested inner flow during spindown. These
conditions are compatible because
they occur at different radii (dynamically important flux is needed
only near the horizon, not at $\astar$) and at different times
(fragmentation early and spindown later). Adopting the spin-evolution of
\citet{Lowell:2023kyu}, $d\chi/d\ln m\simeq-k(\chi-\chi_{\rm eq})$ with
$k\approx9.5$, reaching $\chi_{1}=0.07$ from the post-plunge value
requires $\Delta m_{\rm acc}\approx3$--$5\,\Msun$ for
$\chi_{\rm eq}<0.05$, and hence
$m_{\rm BH,0}\approx15$--$17\,\Msun$ at the time of fragmentation. This
interpretation is internally consistent but remains unverified: it requires the
inner disk to hold a few solar masses after fragmentation, the flow to
reach the magnetically arrested state, and a value of $\chi_{\rm eq}$
that has not been calibrated. The channel predicts
$\chi_{1}\sim0.1$--$0.4$ whenever the spindown phase is absent or
incomplete{ and the plunging fragment is comparable in mass to
\COtwo}; measurements of the primary spin in future GW190814-like
events can therefore test this interpretation.

{The spindown requires the release of rotational energy. The
rotational energy of a BH of mass $m$ and spin $\chi$ is
$E_{\rm rot}=mc^{2}\bigl[1-\sqrt{(1+\sqrt{1-\chi^{2}})/2}\,\bigr]$.
For a $20\,\Msun$ BH, spindown from $\chi\approx0.5$ to $0.07$ releases
$\sim10^{54}\,$erg, far exceeding the $\sim10^{51}$--$10^{52}\,$erg
 collimation-corrected energies of typical LGRB jets. Avoiding such an
 excessive energy release therefore favors a low post-plunge spin,
 $\chi\sim0.1$, for which spindown to $0.07$ releases only $\sim10^{52}\,$erg.

A light plunging fragment produces a modest post-plunge spin. A
fragment of $0.5$--$1\,\Msun$ plunging into a $\approx22\,\Msun$ BH
gives $q\approx0.02$--$0.05$ and $\chi\approx0.07$--$0.14$. This is the
light-plunger limit $\mu<1/2$ derived above for
$m_{\COtwo}=2.6\,\Msun$, in which \COone\ plunges on a nearly radial
orbit producing an even smaller spin.
Thus, a light plunging fragment could account for the low primary spin with little subsequent accretion, so $\mBH\simeq m_{1}$. However, because it widens the orbit of \COtwo\ less, producing the same delay requires fragmentation at a larger birth radius.

}}

\section{Bright-siren Cosmography}
\label{sec:h0}

If SN2019npv is the electromagnetic counterpart preceding GW190814, the
event is a bright standard siren
\citep{Schutz:1986gp,Holz:2005df,MacLeod:2007jd,Chen:2017rfc,Chen:2023dgw}. The
GWs give the luminosity distance along the line of sight to
the counterpart, $\dL = 229.3 \pm 27.1\,\mathrm{Mpc}$ \citep{Singer:2016erz},
while its host galaxy supplies the redshift, $z = 0.055$
(after correcting for the solar system's motion), with the peculiar velocity
marginalized over a Gaussian prior of width
$\sigma_{v} = 300\,\mathrm{km\,s^{-1}}$. Together, they determine $H_{0}$
through the distance--redshift relation
$\dL = \dL(z, H_{0})$ \citep{LIGOScientific:2017adf}.

The inference of $H_{0}$ depends on the assumed redshift prior. We consider
two choices: (i) a volumetric prior,
$p(z)\,dz \propto dV_{c}$, the standard choice in
standard-siren analyses \citep{LIGOScientific:2017adf}; and (ii) a
flat-in-redshift prior, $p(z)=\mathrm{const}$. In both cases, we adopt a
log-uniform prior on the Hubble constant,
$\pi(H_{0})\propto 1/H_{0}$, following
\citet{LIGOScientific:2017adf}, and account for GW selection
effects in the inference.

As shown in Figure~\ref{fig:h0}, under the volumetric prior, GW190814 alone yields
\begin{equation}
H_{0} \;=\; 71.1^{+9.3}_{-7.5}\,\kmsMpc
\end{equation}
(the maximum a posteriori value and minimal $68.3\%$ credible interval),
lying between the Planck~2018 \citep{Planck:2018vyg} and SH0ES
\citep{Riess:2021jrx} values, consistent with each at less than $1\sigma$. The
flat-in-redshift choice raises $H_0$ to
$73.9^{+10.4}_{-8.2}\,\kmsMpc$, closer to the SH0ES measurement.

Figure~\ref{fig:h0} also shows the joint posterior (purple), which combines
GW190814 with GW170817 giving equal weight to both events. The two associations
differ in confidence, however: GW170817's kilonova is an essentially certain
counterpart, whereas the
GW190814--SN2019npv association is more tentative, with $\FAPuncond \approx
0.012$--$0.013$ (Appendix~\ref{sec:host}). A proper population-level analysis will take into account each siren's association confidence, but this is beyond the scope of this paper.

The GW190814 posterior is tighter and more symmetric than GW170817's, which
carries a long tail past $90\,\kmsMpc$, despite having smaller signal-to-noise ratio (SNR). Distance sets $\sim 97\%$ of the
$H_0$ variance and peculiar velocity the rest, and GW190814 is favored on
both counts.

\noindent\textit{Redshift.} Since $H_0 \propto cz/\dL$ and $\sigma_v \approx
300\,\mathrm{km\,s^{-1}}$ is a fixed error on $cz$, peculiar velocity matters
less at higher $z$: $\sim 10\%$ for GW170817 ($cz \approx
3000\,\mathrm{km\,s^{-1}}$) versus $\sim 1.8\%$ for GW190814 ($z = 0.055$).
The peculiar-velocity term that dominates a $z\sim 0.01$ siren is essentially
switched off here.

\noindent\textit{Distance.} The dominant $(2,2)$ mode constrains only the combination
$(1+\cos^{2}\iota)/(2\dL)$, so an edge-on source mimics a more distant face-on
one. GW190814's mass asymmetry ($q \approx 0.11$) excites the $(3,3)$
multipole ($\mathrm{SNR}\approx 6.6$; \citealp{LIGOScientific:2020zkf}), whose
distinct inclination dependence breaks this degeneracy internally. GW170817 is
near-equal-mass, with negligible higher modes, so its distance--inclination
degeneracy is not broken internally, leaving an edge-on tail that maps into
high $H_0$. It was instead broken externally--via very long baseline interferometry (VLBI) astrometry of its
superluminal radio jet \citep{Hotokezaka:2018dfi} and
later afterglow and peculiar-velocity modeling
\citep{Mooley:2022uqa,Palmese:2023beh,Gourdji:2026rqc,DESI:2025rpo}.
GW190814 breaks it intrinsically: with no counterpart but strong
$(3,3)$-mode content, its inclination is constrained within the GW data
alone.

As a future outlook, collapsar sirens may outperform equal-mass BNS kilonovae at the same SNR, because their small $q$ excites higher harmonics that internally break the distance--inclination degeneracy.

\section{Transients from Delayed Supernova-embedded Mergers}
\label{sec:embedded_kn}

\begin{figure*}
    \centering
    \includegraphics[width=\textwidth]{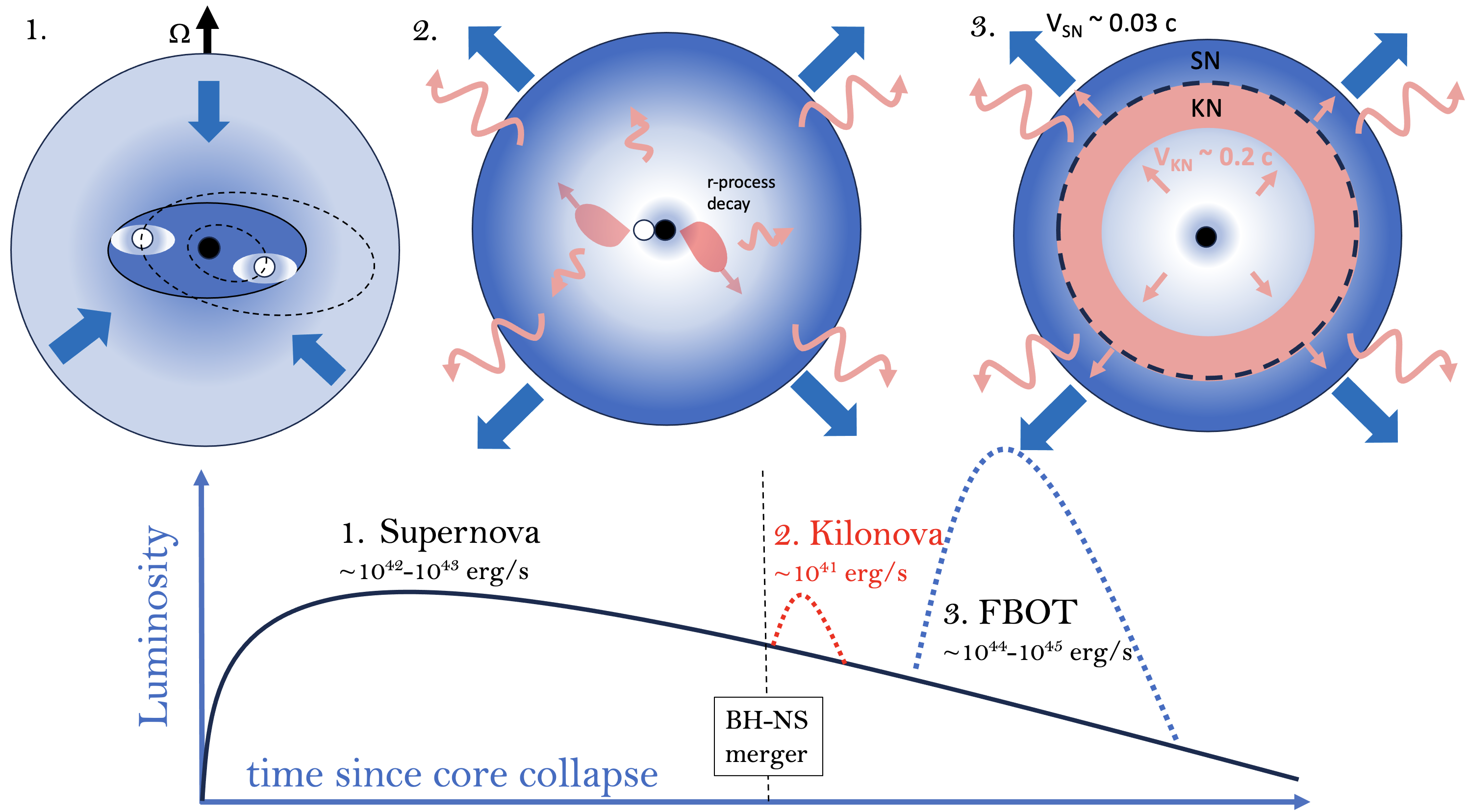}
    \caption{
    Schematic timeline of a delayed compact-object merger embedded inside
    the ejecta of a preceding stripped-envelope supernova.
    \textit{Left:} core collapse forms a central BH surrounded by a
    massive, fragmenting accretion disk.  One or more disk fragments collapse
    into compact objects and undergo gravitational scattering in the potential
    of the central BH.
    \textit{Middle:} after a delay $t_{\rm m}$, one compact object merges with
    the BH inside the expanding supernova ejecta.  If the NS
    is tidally disrupted, the merger launches neutron-rich ejecta of mass
    $M_{\rm KN}$ and velocity $v_{\rm KN}$, whose radioactive $r$-process
    heating can be reprocessed by the overlying supernova material.
    \textit{Right:} the faster merger ejecta catch up with the slower inner
    supernova ejecta, moving at characteristic velocity $v_{\rm SN}$, after a
    time $t_{\rm coll}\simeq v_{\rm SN}t_{\rm m}/(v_{\rm KN}-v_{\rm SN})$.
    Thermalization of the kilonova-ejecta kinetic energy,
    $E_{\rm KN}\simeq M_{\rm KN}v_{\rm KN}^2/2$, can power a luminous
    shock-powered optical flare on a timescale set by the larger of the
    photon diffusion time through the supernova ejecta and the light-crossing
    time of the interaction radius.  Depending on the merger delay and ejecta
    parameters, the observable counterpart may appear as a late-time
    kilonova-powered excess or as an FBOT-like flare near the GW
    event.
    }
    \label{fig:embedded_kn_schematic}
\end{figure*}

Even if GW190814 did not form in a core-collapse environment, this
specific merger was not expected to produce a detectable kilonova. If
the secondary was an NS, the inferred mass ratio and low
primary spin imply a plunging NS--BH merger with negligible tidal
disruption and essentially no remnant disk or ejecta outside the final
BH (e.g., \citealt{Foucart12,GravityCollective:2021kyg}). However, this
conclusion need not apply to future events in the same collapsar-fragment
channel. A lower-mass BH, a rapidly spinning BH, or a
less compact (e.g., subsolar mass; \citealt{Metzger:2024ujc}) collapsar-formed NS could allow tidal disruption
outside the plunge radius, producing neutron-rich ejecta and, hence, a
kilonova.

The distinctive feature of this channel is that the merger occurs inside
the ejecta of the preceding stripped-envelope supernova, for which there have been a few recent candidate events \citep{Kasliwal+25,Hall:2026gov}. If the compact-object
merger is delayed by a time $t_{\rm m}$ after core collapse, the
supernova ejecta have already expanded to radii
$R_{\rm SN} \simeq v_{\rm SN} t_{\rm m}$,
where $v_{\rm SN}$ denotes the characteristic velocity of the inner
supernova ejecta. The merger then launches kilonova ejecta of mass
$M_{\rm KN}$ and velocity $v_{\rm KN}$ into an optically thinning, but
still massive, supernova envelope. Values $M_{\rm KN}\sim10^{-2}$--
$10^{-1}M_\odot$ and $v_{\rm KN}\sim0.1$--$0.3c$ are motivated by
standard kilonova models and by the ejecta inferred for GW170817
(\citealt{Metzger2010,Villar2017}). This geometry produces two possible
electromagnetic signatures near the GW event.

First, the radioactive heating from freshly synthesized $r$-process
material is filtered by the overlying supernova ejecta. For homologously expanding supernova ejecta, the light-curve peaks on a timescale $t_{\rm SN}$ set by when the photon diffusion timescale through the ejecta equals the time since explosion (e.g., \citealt{Arnett1982}).  At later times $t > t_{\rm SN}$, as the optical depth of the ejecta drops, the diffusion timescale obeys
\begin{equation}
    t_{\rm diff}(t) \approx t_{\rm SN}^2/t.
\end{equation}
Thus, a merger-powered radioactive signal can emerge as a distinct
shoulder or rebrightening only once $t_{\rm diff}(t_{\rm m})$ becomes
comparable to or shorter than the intrinsic kilonova timescale
$t_{\rm KN}$. Equivalently,
\begin{equation}
    t_{\rm m} \gtrsim {t_{\rm SN}^2 \over t_{\rm KN}}
    \simeq 80~{\rm d}
    \left({t_{\rm SN}\over 20~{\rm d}}\right)^2
    \left({t_{\rm KN}\over 5~{\rm d}}\right)^{-1}.
\end{equation}
At shorter delays, the kilonova luminosity is trapped and reprocessed
into the broader supernova light curve; at longer delays, it can appear
as a separate transient component close to the time of coalescence.  Heavy $r$-process-rich material from the merger mixed into the supernova ejecta could also modify the ordinary stripped-envelope supernova emission by raising the effective opacity, potentially reddening the optical colors, altering the spectra, and producing a near-infrared excess \citep{BarnesMetzger2022}.

Second, and potentially more luminous, the fast kilonova ejecta can
collide with the slower inner supernova ejecta. The catch-up time after
the merger is
\begin{eqnarray}
    t_{\rm coll} &\simeq& {v_{\rm SN}\over v_{\rm KN}-v_{\rm SN}} t_{\rm m} \nonumber \\
    &
    \simeq& 8~{\rm d}
    \left({v_{\rm SN}\over 5\times10^3~{\rm km~s^{-1}}}\right)
    \left({v_{\rm KN}\over 0.2c}\right)^{-1}
    \left({t_{\rm m}\over 100~{\rm d}}\right),
\end{eqnarray}
where the second equality assumes $v_{\rm KN}\gg v_{\rm SN}$. The
interaction therefore occurs shortly after the GW merger,
rather than near the original supernova maximum. The available kinetic
energy of the kilonova ejecta is
\begin{equation}
    E_{\rm KN} \simeq {1\over 2} M_{\rm KN} v_{\rm KN}^2
    \simeq 2\times10^{51}~{\rm erg}
    \left({M_{\rm KN}\over 5\times10^{-2}M_\odot}\right)
    \left({v_{\rm KN}\over 0.2c}\right)^2.
\end{equation}
This is comparable to the kinetic energy of an ordinary Type Ib
supernova and far exceeds the radiated energy of a normal radioactive
stripped-envelope event. If a fraction $\epsilon_{\rm rad}$ of this
energy is thermalized and escapes on the local diffusion time, the
characteristic luminosity is
\begin{eqnarray}
    &L_{\rm sh}& \sim
    \epsilon_{\rm rad}{E_{\rm KN}\over t_{\rm diff}(t_{\rm m})} \nonumber \\
    &\simeq&
    5\times10^{44}~{\rm erg~s^{-1}}
    \left({\epsilon_{\rm rad}\over 0.1}\right)
    \left({E_{\rm KN}\over 2\times10^{51}~{\rm erg}}\right)
    \left({t_{\rm diff}\over 5~{\rm d}}\right)^{-1}.
\end{eqnarray}
Such luminosities overlap those of fast blue optical transients (FBOTs),
a class of rapidly evolving, luminous, blue events uncovered by modern
wide-field surveys and exemplified by AT2018cow
(\citealt{Drout2014,Perley2019,Ho2019,Margutti2019,Ho2021}), though a couple of these events exhibit hydrogen in their spectra and, hence, are not themselves promising collapsar progenitors.  The
observed duration should be of order
\begin{equation}
    t_{\rm flare} \sim \max\left[t_{\rm diff}(t_{\rm m}), {R_{\rm SN}\over c}\right],
\end{equation}
because the light-crossing time of the interaction radius provides a
minimum geometric smearing time, while photon diffusion through the
overlying supernova ejecta can dominate at earlier merger times.

Thus, collapsar-fragment mergers can produce a qualitatively new class of
``embedded kilonovae'': compact-object merger transients whose radiation
is reprocessed by the supernova ejecta launched weeks to months earlier.
Depending on the merger delay and on the mass, velocity, and opacity of
the merger ejecta, the observable signal may range from a subtle
late-time excess on the supernova light curve to a luminous, shock-powered
FBOT-like flare temporally coincident with the GW event.  These possibilities are illustrated schematically in Figure~\ref{fig:embedded_kn_schematic}.
{An FBOT-like luminosity or timescale alone is insufficient to
distinguish this channel, because collapsars can produce rapidly
evolving luminous emission through several mechanisms. The proposed
diagnostic is the temporal and spatial association of a
stripped-envelope supernova, a GW merger at the same position weeks to
months later, and a subsequent rebrightening.}

A related diagnostic concerns the nature of the secondary itself.  The GW signal alone does not determine whether the $2.6M_\odot$ secondary in GW190814 was an NS or a low-mass BH, a question with key implications for the equation of state of ultrahigh-density matter \citep{Most+20,Fattoyev+20,Godzieba+20,Tsokaros+20,Nathanail+21}.  However, in the collapsar-fragment scenario, the electromagnetic environment may in principle provide additional leverage.  If the secondary were the long-lived NS remnant of an earlier fragment merger, its rotational energy reservoir could be enormous, $E_{\rm rot}\sim 10^{52}-10^{53}\ {\rm erg}$, for millisecond rotation.  Substantial deposition of this energy into the surrounding supernova ejecta would be expected to strongly modify the subsequent transient, for example, by increasing the kinetic energy, broadening the spectral features, or producing a much more luminous engine-powered event.  The apparently ordinary ejecta velocities of SN2019npv implied by the spectral line-widths therefore mildly favor the alternative in which the fragment collapsed promptly to a BH, or in which any NS phase was short-lived or inefficiently coupled to the ejecta.  We emphasize, however, that this inference is conditional on the proposed SN2019npv--GW190814 association and the assumption of at least one previous merger in the disk.

\section{Discussion}
\label{sec:discussion}

{We have explored the possibility that the extreme mass ratio of GW190814 arises because} its $2.6\,\Msun$ secondary formed through
gravitational fragmentation of a neutrino-cooled collapsar disk and
subsequently inspiraled into a central $23\,\Msun$ BH. In this
picture, a single astrophysical environment {can account} for a set of otherwise
disparate features: the extreme mass ratio, a secondary in the putative mass
gap, a slowly spinning primary, a preceding stripped-envelope supernova, and a
candidate bright-siren host.

Statistically, GW190814 is already an outlier: its secondary has a very low
probability of belonging to the standard binary-BH or NS--BH population
\citep{Essick:2021vlx}---a sign not merely of rarity but of a distinct formation pathway. The real difficulty for conventional channels is not
any single parameter, but meeting three at once: an extreme mass ratio, a
mass-gap secondary, and the observed event rate (Figure~\ref{fig:pq}).
The secondary captures this tension. At $2.59\,\Msun$, it sits above
securely measured NSs, unless the equation of state is extremely
stiff or the spin near-maximal \citep{Essick:2020ghc}, yet below the
lightest confirmed BHs \citep{Fishbach:2020ryj,Essick:2020ghc}{,
in an interval that GW230529 has since shown to be sparsely populated
rather than empty \citep{LIGOScientific:2024elc}}.
Collapsar-disk fragmentation resolves this. Because the fragment
forms in situ within the central BH's potential, it need not be sorted into
NS or BH: it may settle into a massive NS, or,
through continued migration and accretion, collapse into a low-mass BH.

{The low spin of the primary ($\chi_{1}<0.07$;
\citealt{LIGOScientific:2020zkf}) is, in this channel, {either a
direct consequence of a light plunging fragment
($m_{\COone}\lesssim1\,\Msun$) or} the outcome of
magnetically arrested accretion of the inner disk after the first
plunge (Section~\ref{sec:nbody}). Whether the flow reaches that state,
and the equilibrium spin it reaches, are the least certain elements
of the scenario; high spins are expected when sustained disk--jet
coupling is absent \citep{Shibata:2023tho,Gottlieb:2025ugy}, and a
moderately spinning primary in a future event of this kind would
indicate an incomplete spindown phase. {The low primary spin and
jet energetics therefore favor a light plunging fragment,
$m_{\COone}\lesssim1\,\Msun$. }} The absence of a jet or LGRB is
unsurprising: GW190814's inclination ($\gtrsim 36^{\circ}$; \citealt{LIGOScientific:2020zkf})
far exceeds the $\lesssim 10^{\circ}$ needed to view an on-axis jet.

The outer-companion merger channel identified in this work predicts that
systems like GW190814 may enter the GW band on nearly circular
orbits, consistent with the observed signal morphology
(Section~\ref{sec:nbody}). In contrast, prompt inner mergers occurring shortly
after formation are expected to retain measurable eccentricity upon entering
the detector band, providing a distinguishing signature of in-disk assembly
\citep{Wu:2026hth}.

More distinctively, \COone's eccentricity is actively reshaped while it is
still scattering off \COtwo. Each pericenter passage of \COtwo\ modifies the
orbit. Unlike GW emission, which monotonically damps
eccentricity, these dynamical encounters can both increase and decrease
eccentricity, producing a nonmonotonic evolution. The resulting waveforms (to
be presented in future work) may therefore deviate from standard eccentric
inspiral templates, which assume smooth GW-driven
circularization, and could potentially evade detection in template-based
searches. This also provides a possible explanation for the absence of a
detected prompt \COone$+$BH merger in GW190814. The merger may have
occurred during detector downtime, within an unfavorable antenna response
region, or through a gas-assisted plunge---occurring while the disk was still
gas-rich---in which rapid orbital decay produced only a limited
GW signal in the sensitive band {(Section~\ref{sec:nbody}). A gas
disk confined to within $\sim30\,\Rg$ of the BH would suffice: gas
torques there would dephase the \COone\ signal relative to vacuum
templates, reducing its detectability in matched-filter searches}.

GW190814 may not be unique but instead part of a broader class of
extreme-mass-ratio mergers. The event GW200210\_092254
($\approx 24+2.8\,\Msun$, $q\approx0.12$) represents a close structural
analog within the GWTC-3 catalog \citep{KAGRA:2021vkt}. However, identifying a preceding stripped-envelope supernova for such
systems is observationally challenging. A supernova comfortably detectable
in wide-field surveys for GW190814 would, at the roughly four-times-larger
distance of GW200210\_092254, fall below the completeness limits of O3-era
transient surveys. This imposes a fundamental observational horizon for testing progenitor-supernova connections, leaving GW190814 as one of the most accessible nearby test cases until next-generation surveys such as Rubin and Roman extend the reach.

Recent subsolar GW candidates---SN2025ulz/S250818k and SN2025adtq/S251112cm---have
been tied to stripped-envelope supernovae and read as products of disk
fragmentation in collapsing massive stars \citep{Kasliwal+25,Hall:2026gov}.
They may be lower-mass, shorter-delay analogues of GW190814, whose more-massive
secondary and longer supernova-to-merger delay simply mark another
point along a continuum within the same collapsar channel.

Because these mergers follow the death of a massive star, they should often be preceded by a stripped-envelope supernova (Type Ib, Ic, Ic-BL, or IIb) and may sometimes produce a long-duration or low-luminosity gamma-ray burst \citep{Piro&Pfahl07,Metzger:2024ujc}. This changes the usual GW follow-up strategy. Rather than only searching for electromagnetic signals after the merger, one of the most informative signatures may occur beforehand. The delay from supernova to merger can span a wide range: from minutes to hours in prompt channels, to months in delayed outer-companion systems. Because the delay can reach months, real-time follow-up should consider
stripped-envelope supernovae from the past several months as candidate
hosts, not just recent weeks \citep{Metzger:2024ujc,Hall:2026gov}. In some cases, the merger may also be followed weeks later by an embedded transient (Section~\ref{sec:embedded_kn}). As a result, archival data from surveys such as the Zwicky Transient Facility (ZTF), ATLAS, Pan-STARRS, and the Dark Energy Survey (DES), together with Fermi and Swift observations, provide a powerful way to identify candidates. SN2019npv could be the first candidate found using this approach. If future GW events show extreme mass ratios, subsolar companions, and pre-merger supernova associations, they would support a common collapsar-disk origin and suggest that GW190814 is the first detected member of a larger population.

\section*{Acknowledgments}
We thank Hsin-Yu Chen for pointing out an issue in our original $H_0$ inference and for helpful discussions.
We also thank Emanuele Berti, Loris Del Grosso, Valerio De Luca, Xander Hall,
Mansi Kasliwal, and Antonella Palmese for helpful comments on the draft.
V.B. acknowledges support from the  NASA Hubble Fellowship grant HST-HF2-51548.001-A awarded by the Space Telescope Science Institute, which is operated by the Association of Universities for Research in Astronomy, Inc., for NASA, under contract NAS5-26555. B.D.M. acknowledges support from NASA (grant 80NSSC26K0299), the National Science Foundation (grant AST-2406637), and the Simons Foundation (727700).  The Flatiron Institute is supported by the Simons Foundation. This work used Delta at the National Center for Supercomputing Applications (NCSA) through allocation PHY260065 from the Advanced Cyberinfrastructure Coordination Ecosystem: Services \& Support (ACCESS) program, which is supported by U.S. National Science Foundation grants
\#2138259, \#2138286, \#2138307, \#2137603, and \#2138296.

\appendix
\twocolumngrid

\section{A Rate Consistent with the Collapsar Channel}
\label{sec:rate}

Beyond the morphological mass-ratio argument, the channel must supply
GW190814-like mergers at the measured rate $R_{\rm GW190814} = 7^{+16}_{-6}\,
\mathrm{Gpc^{-3}\,yr^{-1}}$ \citep{LIGOScientific:2020zkf}, which conventional
formation channels miss by 1--4 orders of magnitude
\citep{Zevin:2020gma,Mandel:2021smh}.

Successful, on-axis long gamma-ray bursts (LGRBs) are the cleanest electromagnetic signature of collapsars \citep{Woosley:1993wj,MacFadyen:1998vz,Woosley:2006fn}, but they likely represent only a subset of engine-forming massive-star
deaths{, and the progenitors of the present channel need not be those of LGRBs (Section~\ref{sec:intro})}. Although LGRBs with spectroscopically confirmed supernovae are associated almost exclusively with broad-lined Type Ic events, the present channel need not require an on-axis, successful relativistic jet, nor a fully stripped progenitor. The essential ingredient is the formation of a rapidly rotating BH surrounded by a massive, fragmentation-prone accretion disk. Such a central engine can plausibly arise in hydrogen-stripped but helium-retaining progenitors --- producing a Type Ib
optical spectrum --- especially if the jet is choked, misaligned, weak, or fails to break out of the stellar envelope \citep{Bromberg:2011wb}. The Type Ib classification of SN2019npv is therefore not the canonical LGRB-SN outcome, but it is not obviously inconsistent with a collapsar-like central engine
either. We note also that no LGRB was seen from GW190814 itself, whose inclination ($\gtrsim 36^{\circ}$) far exceeds the $\lesssim 10^{\circ}$ viewing angle an on-axis jet would require.

With these caveats in mind, we take the intrinsic LGRB rate as a fiducial collapsar formation rate and provide order-of-magnitude estimate:
\begin{equation}
R_{\rm GW190814} = f_{m}\, {\epsilon_{\rm frag}\,\epsilon_{\rm LGRB}\,} R_{\rm LGRB},
\label{eq:rate}
\end{equation}
where $f_{m}$ is the fraction of collapsar disks whose central remnant is consistent with $m_1\sim 23\,M_{\odot}$ of GW190814. This fraction depends on
highly uncertain core-collapse dynamics, progenitor properties, metallicity, and the initial
mass function, but it is plausibly in the $1\%$--$10\%$ range.\footnote{Based on rough estimates from the \citet{Fryer:2011cx} remnant-mass prescription as implemented in the COMPAS population synthesis code \citep{COMPASTeam:2025fjl}, with small $f_{m}$ favored at high metallicity and large $f_{m}$ at low metallicity.}
{The two remaining factors have opposing effects on the inferred rate.
$\epsilon_{\rm LGRB}\equiv R_{\rm coll}/R_{\rm LGRB}>1$ is the number
of collapsars per successful on-axis burst: choked or failed jets
\citep{Bromberg:2011wb} and the more numerous low-luminosity bursts
\citep{Liang:2006ci,Guetta:2006gq} imply a plausible value of order
$10$ or larger. $\epsilon_{\rm frag}<1$ is the fraction of collapsar disks in the
relevant central-mass range that fragment, cool, and form compact objects.
Neither factor is known; the observed rate constrains only their
product, $\epsilon\equiv\epsilon_{\rm frag}\,\epsilon_{\rm LGRB}$.}

\begin{figure}[!htbp]
\centering
\includegraphics[width=\linewidth]{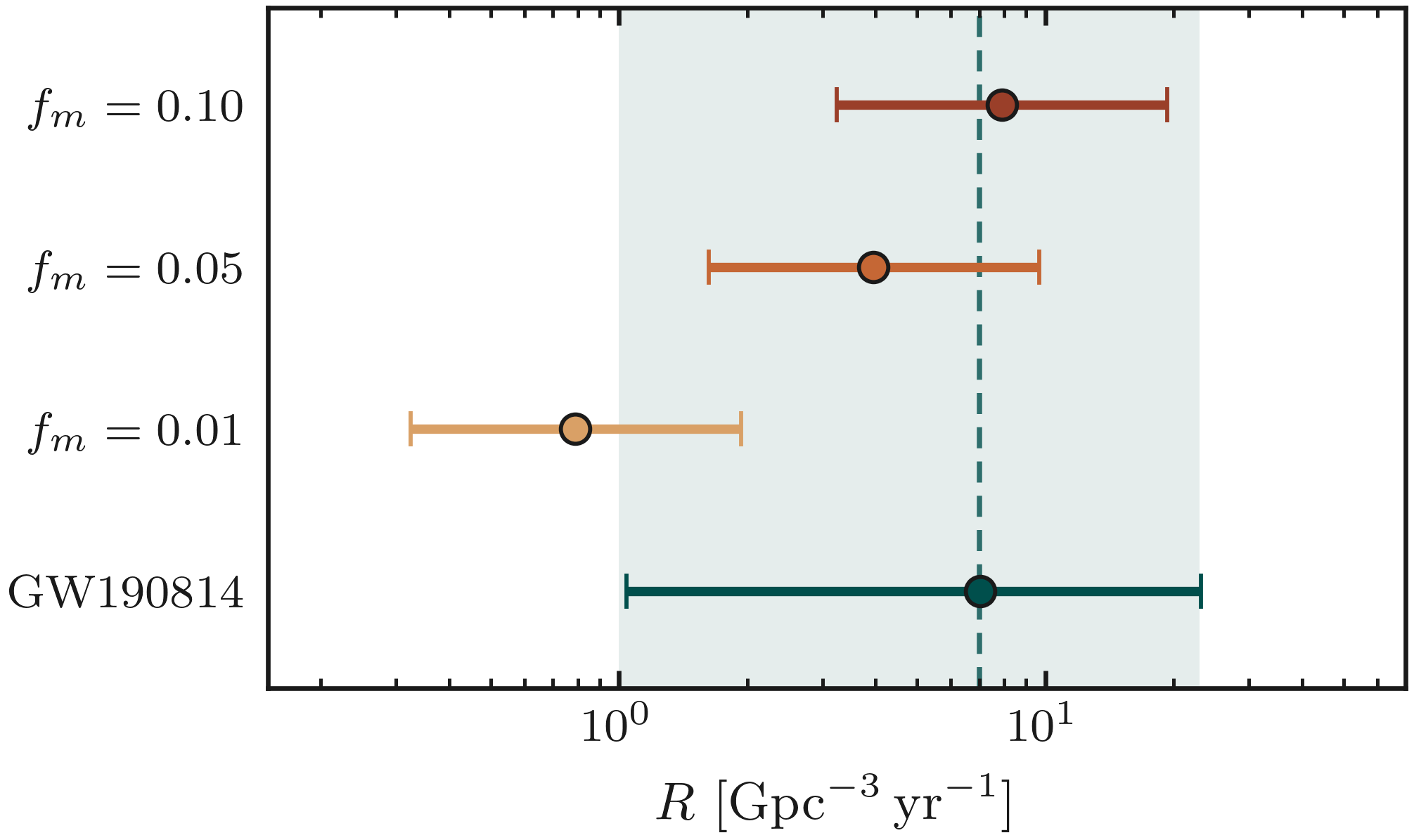}
\caption{Predicted local rate of GW190814-like mergers,
$R_{\rm GW190814} = f_{m}\,R_{\rm LGRB}$ (Equation~\ref{eq:rate} {with $\epsilon_{\rm frag}\epsilon_{\rm LGRB}=1$}), for  $f_{m} = 0.01$, $0.05$, and $0.10$ (top three rows), compared with
the measured LVK rate for GW190814 (bottom). The shaded band marks
the measured $90\%$ interval ($1$--$23\,\mathrm{Gpc^{-3}\,yr^{-1}}$) and the
dashed line its median ($7\,\mathrm{Gpc^{-3}\,yr^{-1}}$;
\citealp{LIGOScientific:2020zkf}); predicted rates use
$R_{\rm LGRB} = 79^{+57}_{-33}\,\mathrm{Gpc^{-3}\,yr^{-1}}$
\citep{Ghirlanda:2022edk}. The $f_{m}=0.05$ and $f_{m}=0.10$ cases overlap the
measured posterior, while $f_{m}=0.01$ lies at the low-rate edge.}
\label{fig:rate}
\end{figure}

With the beaming-corrected local rate $R_{\rm LGRB} = 79^{+57}_{-33}\,
\mathrm{Gpc^{-3}\,yr^{-1}}$ \citep{Ghirlanda:2022edk}, Equation~\ref{eq:rate}
predicts $0.8$, $3.9$, and $7.9\,\mathrm{Gpc^{-3}\,yr^{-1}}$ for
$f_{m} = 0.01,\,0.05,\,0.10$ {and $\epsilon=1$}. The $f_{m}=0.05$ and $f_{m}=0.10$ cases fall
squarely inside the measured posterior, while $f_{m}=0.01$ sits at its low-rate
edge (Figure~\ref{fig:rate}).
{Across $f_{m}=0.01$--$0.1$, the measured $90\%$ interval
($1$--$23\,\mathrm{Gpc^{-3}\,yr^{-1}}$) is reproduced for
$\epsilon\approx0.1$--$30$.
Agreement with the current measurement therefore establishes
consistency, but several GW190814-like events, with and without supernova counterparts,
are required to test the predicted rate.}

\section{Host Association and Chance Coincidence}
\label{sec:host}
Could the SN2019npv host lie inside the GW190814 credible volume by
chance? Here we present a simplified estimate following the host-association methodology of \citet{Hall:2026gov}. One can
measure 3D spatial agreement between two probability
distributions,
\begin{equation}
\mathcal{I} \;=\; \int \pGW(\mathbf{x})\,\pEM(\mathbf{x})\,d^{3}\mathbf{x},
\label{eq:overlap}
\end{equation}
where $\pGW$ is the GW posterior over comoving position $\mathbf{x}$,
and $\pEM$ is the electromagnetic (EM) source density at angular position
WISEA J005332.35$-$234955.8 with distance set by $z = 0.055 \pm
0.001$, propagated to comoving distance. Because this propagation
depends on $H_{0}$, we use both the
Planck~2018 \citep{Planck:2018vyg} and SH0ES \citep{Riess:2021jrx}
cosmologies, which bracket the current $H_{0}$ tension and set the
reported ranges. $\mathcal{I} \gg 1$ means the
GW posterior is peaked at the EM source; $\mathcal{I} \ll 1$ that
it excludes it. For GW190814~$\times$~SN2019npv, we find
\begin{equation}
\log_{10} \mathcal{I} \;\simeq\; 3.64\text{--}3.80,
\label{eq:Iobs}
\end{equation}
where the lower (upper) end corresponds to the Planck~2018 (SH0ES)
cosmology. Either way, the GW posterior is more than 3 orders of
magnitude more concentrated at the SN2019npv position than it would
be if spread uniformly.

We interpret $\log_{10}\mathcal{I}_{\mathrm{obs}}$ by comparing it
to simulations from two scenarios: a random, unrelated EM source
($H_{0}$) and a true GW counterpart ($H_{\mathrm{assoc}}$). The
first uses uniformly distributed 3D positions, while the second
uses positions drawn from the GW posterior.

A chance association requires two things to occur jointly: a Type
Ib supernova must appear inside the GW credible volume during the
time window of interest, \emph{and} its 3D position
inside that volume must match the GW posterior at least as well as
the observed value.

The first piece is a Poisson counting calculation. If $\Rsn$ is the
local Type Ib volumetric rate and $V_{90}$, the GW 90\% credible
volume, the expected number of unrelated Type Ib supernovae inside
$V_{90}$ during $\Delta t$ is
\begin{equation}
\lambda \;=\; \Rsn\, V_{90}\, \Delta t,
\label{eq:lambda_poisson}
\end{equation}
and the probability that at least one such chance supernova appears
is
\begin{equation}
\Pcoinc \;=\; 1 - e^{-\lambda} \;\approx\; 3.7\%\text{--}4.6\%,
\label{eq:pcoinc}
\end{equation}
using $V_{90} = 3.21 \times 10^{4}\,\mathrm{Mpc^{3}}$ (comoving), a fiducial
Type Ib rate
$\Rsn = 0.8\times 10^{-5}\,\mathrm{Mpc^{-3}\,yr^{-1}}\, h^3_{70}$
\citep{Pessi:2025wht}, and $\Delta t = 60\,\mathrm{d}$ matching the
SN2019npv pre-trigger gap. Because the rate carries an $h^3_{70}$
factor and the credible volume depends on cosmology, $\Pcoinc$
spans the two $H_0$ choices. Even with no physical association, the
probability that \emph{some} Type Ib supernova lands in the
credible volume within 60 days is about $3.7\%$ (Planck~2018) to
$4.6\%$ (SH0ES).

$\Pcoinc$ measures how often the credible volume is populated by
some unrelated supernova; it does not address how well the position
of the supernova we actually found matches the GW posterior. To
test that, we condition on positions already inside the credible
volume and ask whether the observed $\log_{10}\mathcal{I}$ is
exceeded by random within-volume draws. Drawing uniformly inside
the 90\% 3D credible region and computing $\log_{10}\mathcal{I}$
for each, the fraction at or above the observed value is the
conditional false-alarm probability ($\FAPcond$) $\approx 0.26$--$0.36$
(SH0ES--Planck~2018): SN2019npv sits comfortably inside the
credible cylinder but not at its global maximum.

The full probability that an unrelated Type Ib supernova lands in
the credible volume \emph{and} matches the GW posterior as well as
SN2019npv does is therefore
\begin{equation}
\FAPuncond \;=\; \Pcoinc \times \FAPcond \;\approx\; 0.012\text{--}0.013.
\label{eq:fap_uncond}
\end{equation}
This is the total probability that the GW190814--SN2019npv
coincidence arises by chance, given the Type Ib rate, the credible
volume, and the time window. Expressed as a Gaussian significance,
$\FAPuncond$ corresponds to $2.21\sigma$--$2.25\sigma$ (a one-sided
$2.25\sigma$ under SH0ES), so the single GW190814~$\times$~SN2019npv
coincidence currently sits at $\sim\!2.2\sigma$, below the $3\sigma$
threshold.

\paragraph{Likelihood ratios.}
The true-positive rate (TPR) at the observed threshold, measured by
injecting synthetic associations and counting the recovered
fraction, is $\mathrm{TPR} \approx 0.29$--$0.37$. Dividing by each
FAP gives two likelihood ratios:
\begin{align}
\Lambda_{\mathrm{uncond}} &= \mathrm{TPR}/\FAPuncond \approx 22\text{--}30, \\
\Lambda_{\mathrm{cond}}   &= \mathrm{TPR}/\FAPcond   \approx 0.8\text{--}1.4.
\label{eq:lambdas}
\end{align}
$\Lambda_{\mathrm{uncond}} \approx 22$--$30$ indicates that the
joint observation --- a Type Ib supernova inside the GW credible
volume at the right redshift --- is roughly 1--1.5
orders of magnitude more probable under physical association than
under a random full-sky chance match. $\Lambda_{\mathrm{cond}}
\approx 0.8$--$1.4$ straddles unity, indicating that SN2019npv's
specific position within the credible volume is unremarkable: the
host is well inside the volume but not at its global maximum. (Both
ranges span the Planck~2018 and SH0ES cosmologies, with SH0ES
favoring the higher, more association-leaning end.)

For $N$ independent GW~$\times$~EM coincidences each as well
aligned as SN2019npv, the joint chance probability is approximately
\begin{equation}
P(C_{N}) \;\approx\; \FAPuncond^{N}.
\end{equation}
For $\FAPuncond \approx 0.012$--$0.013$, the present single event sits
at $\sim\!2.2\sigma$, so just one more comparable coincidence
($N = 2$) would cross the $3\sigma$ threshold, while reaching
$5\sigma$ requires $N \approx 4$ such events, provided they are
genuinely independent and each individually as well aligned as
SN2019npv.

\bibliographystyle{aasjournal}
\bibliography{refs}

\begin{thebibliography}{}
\expandafter\ifx\csname natexlab\endcsname\relax\def\natexlab#1{#1}\fi
\providecommand{\url}[1]{\href{#1}{#1}}
\providecommand{\dodoi}[1]{doi:~\href{http://doi.org/#1}{\nolinkurl{#1}}}
\providecommand{\doeprint}[1]{\href{http://ascl.net/#1}{\nolinkurl{http://ascl.net/#1}}}
\providecommand{\doarXiv}[1]{\href{https://arxiv.org/abs/#1}{\nolinkurl{https://arxiv.org/abs/#1}}}

\bibitem[{Abac {et~al.}(2024)}]{LIGOScientific:2024elc}
Abac, A.~G., {et~al.} 2024, ApJL, 970, L34, \dodoi{10.3847/2041-8213/ad5beb}

\bibitem[{{Abbott} {et~al.}(2017)}]{LIGOScientific:2017adf}
{Abbott}, B.~P., {et~al.} 2017, Natur, 551, 85, \dodoi{10.1038/nature24471}

\bibitem[{{Abbott} {et~al.}(2020)}]{LIGOScientific:2020zkf}
{Abbott}, R., {et~al.} 2020, ApJL, 896, L44, \dodoi{10.3847/2041-8213/ab960f}

\bibitem[{Abbott {et~al.}(2023)}]{KAGRA:2021vkt}
Abbott, R., {et~al.} 2023, PhRvX, 13, 041039,
  \dodoi{10.1103/PhysRevX.13.041039}

\bibitem[{{Ackley} {et~al.}(2020)}]{Ackley:2020qkz}
{Ackley}, K., {et~al.} 2020, A\&A, 643, A113,
  \dodoi{10.1051/0004-6361/202037669}

\bibitem[{{Aghanim} {et~al.}(2020)}]{Planck:2018vyg}
{Aghanim}, N., {et~al.} 2020, A\&A, 641, A6,
  \dodoi{10.1051/0004-6361/201833910}

\bibitem[{Amsellem {et~al.}(2026)}]{DESI:2025rpo}
Amsellem, A.~J., {et~al.} 2026, ApJ, 1001, 157,
  \dodoi{10.3847/1538-4357/ae4b37}

\bibitem[{{Andreoni} {et~al.}(2020)}]{Andreoni:2019qgh}
{Andreoni}, I., {et~al.} 2020, ApJ, 890, 131, \dodoi{10.3847/1538-4357/ab6a1b}

\bibitem[{{Antoniadis} {et~al.}(2022){Antoniadis}, {Aguilera-Dena},
  {Vigna-G{\'o}mez}, {Kramer}, {Langer}, {M{\"u}ller}, {Tauris}, {Wang}, \&
  {Xu}}]{Antoniadis:2021dhe}
{Antoniadis}, J., {Aguilera-Dena}, D.~R., {Vigna-G{\'o}mez}, A., {et~al.} 2022,
  A\&A, 657, L6, \dodoi{10.1051/0004-6361/202142322}

\bibitem[{{Arnett}(1982)}]{Arnett1982}
{Arnett}, W.~D. 1982, ApJ, 253, 785, \dodoi{10.1086/159681}

\bibitem[{Baibhav {et~al.}(2026)Baibhav, De~Luca, Del~Grosso, Berti, Metzger,
  \& Hui}]{Baibhav:2026iny}
Baibhav, V., De~Luca, V., Del~Grosso, L., {et~al.} 2026.
\newblock \doarXiv{2609.25225}

\bibitem[{{Barnes} \& {Metzger}(2022)}]{BarnesMetzger2022}
{Barnes}, J., \& {Metzger}, B.~D. 2022, \apjl, 939, L29,
  \dodoi{10.3847/2041-8213/ac9b41}

\bibitem[{{Bianco} {et~al.}(2014){Bianco}, {Modjaz}, {Hicken}, {Friedman},
  {Kirshner}, {Bloom}, {Challis}, {Marion}, \& {Wood-Vasey}}]{Bianco:2014mna}
{Bianco}, F.~B., {Modjaz}, M., {Hicken}, M., {et~al.} 2014, ApJS, 213, 19,
  \dodoi{10.1088/0067-0049/213/2/19}

\bibitem[{Blandford \& Znajek(1977)}]{Blandford:1977ds}
Blandford, R.~D., \& Znajek, R.~L. 1977, MNRAS, 179, 433,
  \dodoi{10.1093/mnras/179.3.433}

\bibitem[{{Broekgaarden} {et~al.}(2022){Broekgaarden}, {Stevenson}, \&
  {Thrane}}]{Broekgaarden:2022nst}
{Broekgaarden}, F.~S., {Stevenson}, S., \& {Thrane}, E. 2022, ApJ, 938, 45,
  \dodoi{10.3847/1538-4357/ac8879}

\bibitem[{{Bromberg} {et~al.}(2012){Bromberg}, {Nakar}, {Piran}, \&
  {Sari}}]{Bromberg:2011wb}
{Bromberg}, O., {Nakar}, E., {Piran}, T., \& {Sari}, R. 2012, ApJ, 749, 110,
  \dodoi{10.1088/0004-637X/749/2/110}

\bibitem[{{Chambers} {et~al.}(1996){Chambers}, {Wetherill}, \&
  {Boss}}]{1996Icar..119..261C}
{Chambers}, J.~E., {Wetherill}, G.~W., \& {Boss}, A.~P. 1996, Icar, 119, 261,
  \dodoi{10.1006/icar.1996.0019}

\bibitem[{{Chatterjee} {et~al.}(2008){Chatterjee}, {Ford}, {Matsumura}, \&
  {Rasio}}]{Chatterjee:2007ed}
{Chatterjee}, S., {Ford}, E.~B., {Matsumura}, S., \& {Rasio}, F.~A. 2008, ApJ,
  686, 580, \dodoi{10.1086/590227}

\bibitem[{{Chen} {et~al.}(2018){Chen}, {Fishbach}, \& {Holz}}]{Chen:2017rfc}
{Chen}, H.-Y., {Fishbach}, M., \& {Holz}, D.~E. 2018, Natur, 562, 545,
  \dodoi{10.1038/s41586-018-0606-0}

\bibitem[{Chen {et~al.}(2024)Chen, Talbot, \& Chase}]{Chen:2023dgw}
Chen, H.-Y., Talbot, C., \& Chase, E.~A. 2024, PhRvL, 132, 191003,
  \dodoi{10.1103/PhysRevLett.132.191003}

\bibitem[{Chen \& Beloborodov(2007)}]{Chen:2006rra}
Chen, W.-X., \& Beloborodov, A.~M. 2007, ApJ, 657, 383, \dodoi{10.1086/508923}

\bibitem[{{Chen} \& {Metzger}(2025)}]{Chen:2025uwd}
{Chen}, Y.-X., \& {Metzger}, B.~D. 2025, ApJL, 991, L22,
  \dodoi{10.3847/2041-8213/ae045d}

\bibitem[{{Drout} {et~al.}(2011){Drout}, {Soderberg}, {Gal-Yam}, {Cenko},
  {Fox}, {Leonard}, {Sand}, {Moon}, {Arcavi}, \& {Green}}]{Drout:2010ww}
{Drout}, M.~R., {Soderberg}, A.~M., {Gal-Yam}, A., {et~al.} 2011, ApJ, 741, 97,
  \dodoi{10.1088/0004-637X/741/2/97}

\bibitem[{{Drout} {et~al.}(2014){Drout}, {Chornock}, {Soderberg}, {Sanders},
  {McKinnon}, {Rest}, {Foley}, {Milisavljevic}, {Margutti}, {Berger},
  {Calkins}, {Fong}, {Gezari}, {Huber}, {Kankare}, {Kirshner}, {Leibler},
  {Lunnan}, {Mattila}, {Narayan}, {Riess}, {Roth}, {Scolnic}, {Smartt},
  {Tonry}, {Burgett}, {Chambers}, {Flewelling}, {Hodapp}, {Kaiser}, {Magnier},
  {Wainscoat}, \& {Waters}}]{Drout2014}
{Drout}, M.~R., {Chornock}, R., {Soderberg}, A.~M., {et~al.} 2014, ApJ, 794,
  23, \dodoi{10.1088/0004-637X/794/1/23}

\bibitem[{{Essick} {et~al.}(2022){Essick}, {Farah}, {Galaudage}, {Talbot},
  {Fishbach}, {Thrane}, \& {Holz}}]{Essick:2021vlx}
{Essick}, R., {Farah}, A., {Galaudage}, S., {et~al.} 2022, ApJ, 926, 34,
  \dodoi{10.3847/1538-4357/ac3978}

\bibitem[{Essick \& Landry(2020)}]{Essick:2020ghc}
Essick, R., \& Landry, P. 2020, ApJ, 904, 80, \dodoi{10.3847/1538-4357/abbd3b}

\bibitem[{Fattoyev {et~al.}(2020)Fattoyev, Horowitz, Piekarewicz, \&
  Reed}]{Fattoyev+20}
Fattoyev, F.~J., Horowitz, C.~J., Piekarewicz, J., \& Reed, B. 2020, PhRvC,
  102, 065805, \dodoi{10.1103/PhysRevC.102.065805}

\bibitem[{Fishbach {et~al.}(2020)Fishbach, Essick, \& Holz}]{Fishbach:2020ryj}
Fishbach, M., Essick, R., \& Holz, D.~E. 2020, ApJL, 899, L8,
  \dodoi{10.3847/2041-8213/aba7b6}

\bibitem[{{Ford} \& {Rasio}(2008)}]{Ford:2007ea}
{Ford}, E.~B., \& {Rasio}, F.~A. 2008, ApJ, 686, 621, \dodoi{10.1086/590926}

\bibitem[{{Foucart}(2012)}]{Foucart12}
{Foucart}, F. 2012, \prd, 86, 124007, \dodoi{10.1103/PhysRevD.86.124007}

\bibitem[{{Fryer} {et~al.}(2012){Fryer}, {Belczynski}, {Wiktorowicz},
  {Dominik}, {Kalogera}, \& {Holz}}]{Fryer:2011cx}
{Fryer}, C.~L., {Belczynski}, K., {Wiktorowicz}, G., {et~al.} 2012, ApJ, 749,
  91, \dodoi{10.1088/0004-637X/749/1/91}

\bibitem[{{Ghirlanda} \& {Salvaterra}(2022)}]{Ghirlanda:2022edk}
{Ghirlanda}, G., \& {Salvaterra}, R. 2022, ApJ, 932, 10,
  \dodoi{10.3847/1538-4357/ac6e43}

\bibitem[{Godzieba {et~al.}(2021)Godzieba, Radice, \& Bernuzzi}]{Godzieba+20}
Godzieba, D.~A., Radice, D., \& Bernuzzi, S. 2021, ApJ, 908, 122,
  \dodoi{10.3847/1538-4357/abd4dd}

\bibitem[{Gottlieb {et~al.}(2023)Gottlieb, Jacquemin-Ide, Lowell, Tchekhovskoy,
  \& Ramirez-Ruiz}]{Gottlieb:2023cgm}
Gottlieb, O., Jacquemin-Ide, J., Lowell, B., Tchekhovskoy, A., \& Ramirez-Ruiz,
  E. 2023, ApJL, 952, L32, \dodoi{10.3847/2041-8213/ace779}

\bibitem[{Gottlieb {et~al.}(2025)Gottlieb, Metzger, Issa, Li, Renzo, \&
  Isi}]{Gottlieb:2025ugy}
Gottlieb, O., Metzger, B.~D., Issa, D., {et~al.} 2025, ApJL, 993, L54,
  \dodoi{10.3847/2041-8213/ae0d81}

\bibitem[{Gourdji {et~al.}(2026)Gourdji, Deller, Flynn, Govreen-Segal, Howlett,
  Mooley, \& Nakar}]{Gourdji:2026rqc}
Gourdji, K., Deller, A.~T., Flynn, C., {et~al.} 2026, ApJ, 1005, 93,
  \dodoi{10.3847/1538-4357/ae706c}

\bibitem[{{Guetta} \& {Della Valle}(2007)}]{Guetta:2006gq}
{Guetta}, D., \& {Della Valle}, M. 2007, ApJL, 657, L73, \dodoi{10.1086/511417}

\bibitem[{{Hall} {et~al.}(2026){Hall}, {Ahumada}, {Gassert}, {Palmese},
  {Metzger}, {Kasliwal}, {Bulla}, {Gruen}, {Stein}, {Fremling}, {Anand},
  {Andreoni}, {Busmann}, {Cabrera}, {Christinzio}, {Freeburn}, {Maga{\~n}a
  Hernandez}, {Hu}, {O'Connor}, {Jiang}, {Liu}, {Zhao}, {Bellm}, {Cook},
  {Coughlin}, {Dekany}, {Graham}, \& {Laher}}]{Hall:2026gov}
{Hall}, X.~J., {Ahumada}, T., {Gassert}, J., {et~al.} 2026, arXiv e-prints,
  arXiv:2605.10940, \dodoi{10.48550/arXiv.2605.10940}

\bibitem[{{Ho} {et~al.}(2019){Ho}, {Phinney}, {Kulkarni}, {Dong}, {De},
  {Corsi}, {Hallinan}, {Mooley}, {Nakar}, {Bellm}, {Burdge}, {Cenko}, {Dobie},
  {Frail}, {Kasliwal}, {Murphy}, {Ngeow}, {Nugent}, {Ofek}, {Perley},
  {Petitpas}, {Riddle}, \& {Soderberg}}]{Ho2019}
{Ho}, A.~Y.~Q., {Phinney}, E.~S., {Kulkarni}, S.~R., {et~al.} 2019, ApJ, 871,
  73, \dodoi{10.3847/1538-4357/aaf473}

\bibitem[{{Ho} {et~al.}(2021){Ho}, {Perley}, {Gal-Yam}, {Lunnan}, {Sollerman},
  {Schulze}, {Das}, {Dobie}, {Yao}, {Fremling}, {Adams}, {Anand}, {Andreoni},
  {Bellm}, {Bruch}, {Burdge}, {Castro-Tirado}, {De}, {Dekany}, {Drake}, {Duev},
  {Graham}, {Helou}, {Kaplan}, {Kasliwal}, {Kool}, {Kulkarni}, {Mahabal},
  {Medford}, {Miller}, {Nordin}, {Ofek}, {Petitpas}, {Riddle}, {Sharma},
  {Smith}, {Stewart}, {Tartaglia}, \& {Tzanidakis}}]{Ho2021}
{Ho}, A.~Y.~Q., {Perley}, D.~A., {Gal-Yam}, A., {et~al.} 2021, ApJ, 920, 35,
  \dodoi{10.3847/1538-4357/ac0fcb}

\bibitem[{{Holz} \& {Hughes}(2005)}]{Holz:2005df}
{Holz}, D.~E., \& {Hughes}, S.~A. 2005, ApJ, 629, 15, \dodoi{10.1086/431341}

\bibitem[{Hotokezaka {et~al.}(2019)Hotokezaka, Nakar, Gottlieb, Nissanke,
  Masuda, Hallinan, Mooley, \& Deller}]{Hotokezaka:2018dfi}
Hotokezaka, K., Nakar, E., Gottlieb, O., {et~al.} 2019, NatAs, 3, 940,
  \dodoi{10.1038/s41550-019-0820-1}

\bibitem[{Issa {et~al.}(2026)Issa, Lowell, Jacquemin-Ide, Liska, \&
  Tchekhovskoy}]{Issa:2025jzq}
Issa, D., Lowell, B., Jacquemin-Ide, J., Liska, M., \& Tchekhovskoy, A. 2026,
  PhRvD, 113, 083020, \dodoi{10.1103/fzhd-hcpp}

\bibitem[{Jacquemin-Ide {et~al.}(2024)Jacquemin-Ide, Gottlieb, Lowell, \&
  Tchekhovskoy}]{Jacquemin-Ide:2023aax}
Jacquemin-Ide, J., Gottlieb, O., Lowell, B., \& Tchekhovskoy, A. 2024, ApJ,
  961, 212, \dodoi{10.3847/1538-4357/ad02f0}

\bibitem[{{Juri{\'c}} \& {Tremaine}(2008)}]{Juric:2007dx}
{Juri{\'c}}, M., \& {Tremaine}, S. 2008, ApJ, 686, 603, \dodoi{10.1086/590047}

\bibitem[{{Kasliwal} {et~al.}(2025){Kasliwal}, {Ahumada}, {Stein},
  {Karambelkar}, {Hall}, {Singh}, {Fremling}, {Metzger},
  {et~al.}}]{Kasliwal+25}
{Kasliwal}, M.~M., {Ahumada}, T., {Stein}, R., {et~al.} 2025, \apjl, 995, L59,
  \dodoi{10.3847/2041-8213/ae2000}

\bibitem[{{Kilpatrick} {et~al.}(2021)}]{GravityCollective:2021kyg}
{Kilpatrick}, C.~D., {et~al.} 2021, ApJ, 923, 258,
  \dodoi{10.3847/1538-4357/ac23c6}

\bibitem[{{Laskar} \& {Petit}(2017)}]{2017A&A...605A..72L}
{Laskar}, J., \& {Petit}, A.~C. 2017, A\&A, 605, A72,
  \dodoi{10.1051/0004-6361/201630022}

\bibitem[{{Lerner} {et~al.}(2026){Lerner}, {Stone}, \& {Ofengeim}}]{Lerner+26}
{Lerner}, Y., {Stone}, N.~C., \& {Ofengeim}, D.~D. 2026, \mnras, 545, staf1835,
  \dodoi{10.1093/mnras/staf1835}

\bibitem[{{Liang} {et~al.}(2007){Liang}, {Zhang}, {Virgili}, \&
  {Dai}}]{Liang:2006ci}
{Liang}, E., {Zhang}, B., {Virgili}, F., \& {Dai}, Z.~G. 2007, ApJ, 662, 1111,
  \dodoi{10.1086/517959}

\bibitem[{Lowell {et~al.}(2025)Lowell, Jacquemin-Ide, Liska, \&
  Tchekhovskoy}]{Lowell:2025prd}
Lowell, B., Jacquemin-Ide, J., Liska, M., \& Tchekhovskoy, A. 2025, PhRvD, 112,
  123023.
\newblock \doarXiv{2502.17559}

\bibitem[{Lowell {et~al.}(2024)Lowell, Jacquemin-Ide, Tchekhovskoy, \&
  Duncan}]{Lowell:2023kyu}
Lowell, B., Jacquemin-Ide, J., Tchekhovskoy, A., \& Duncan, A. 2024, ApJ, 960,
  82, \dodoi{10.3847/1538-4357/ad09af}

\bibitem[{Lu {et~al.}(2021)Lu, Beniamini, \& Bonnerot}]{Lu:2020gfh}
Lu, W., Beniamini, P., \& Bonnerot, C. 2021, MNRAS, 500, 1817,
  \dodoi{10.1093/mnras/staa3372}

\bibitem[{{MacFadyen} \& {Woosley}(1999)}]{MacFadyen:1998vz}
{MacFadyen}, A., \& {Woosley}, S.~E. 1999, ApJ, 524, 262,
  \dodoi{10.1086/307790}

\bibitem[{{MacLeod} \& {Hogan}(2008)}]{MacLeod:2007jd}
{MacLeod}, C.~L., \& {Hogan}, C.~J. 2008, PhRvD, 77, 043512,
  \dodoi{10.1103/PhysRevD.77.043512}

\bibitem[{{Maga{\~n}a Hernandez} \& {Breivik}(2025)}]{MaganaHernandez:2025sra}
{Maga{\~n}a Hernandez}, I., \& {Breivik}, K. 2025, arXiv e-prints.
\newblock \doarXiv{2511.16648}

\bibitem[{{Mandel} \& {Broekgaarden}(2022)}]{Mandel:2021smh}
{Mandel}, I., \& {Broekgaarden}, F.~S. 2022, LRR, 25, 1,
  \dodoi{10.1007/s41114-021-00034-3}

\bibitem[{{Mandel} \& {de Mink}(2016)}]{Mandel:2015qlu}
{Mandel}, I., \& {de Mink}, S.~E. 2016, MNRAS, 458, 2634,
  \dodoi{10.1093/mnras/stw379}

\bibitem[{Mandel {et~al.}(2025)}]{COMPASTeam:2025fjl}
Mandel, I., {et~al.} 2025, ApJS, 280, 43, \dodoi{10.3847/1538-4365/adf8d0}

\bibitem[{{Marchant} {et~al.}(2016){Marchant}, {Langer}, {Podsiadlowski},
  {Tauris}, \& {Moriya}}]{Marchant:2016wow}
{Marchant}, P., {Langer}, N., {Podsiadlowski}, P., {Tauris}, T.~M., \&
  {Moriya}, T.~J. 2016, A\&A, 588, A50, \dodoi{10.1051/0004-6361/201628133}

\bibitem[{{Marchant} {et~al.}(2021){Marchant}, {Pappas}, {Gallegos-Garcia},
  {Berry}, {Taam}, {Kalogera}, \& {Podsiadlowski}}]{Marchant:2021hiv}
{Marchant}, P., {Pappas}, K. M.~W., {Gallegos-Garcia}, M., {et~al.} 2021, A\&A,
  650, A107, \dodoi{10.1051/0004-6361/202039992}

\bibitem[{{Margutti} {et~al.}(2019){Margutti}, {Metzger}, {Chornock}, {Vurm},
  {Roth}, {Grefenstette}, {Savchenko}, {Margalit}, {Migliori}, {Milisavljevic},
  {Alexander}, {Bietenholz}, {Blanchard}, {Boehm}, {Brethauer}, {Chilingarian},
  {Coppejans}, {Dimitriadis}, {Fong}, {Gomboc}, {Leja}, {Lunnan}, {Patnaude},
  {Terreran}, {Berger}, {Cargile}, {Challis}, {Chauhan}, {Cowperthwaite},
  {Eftekhari}, {Foley}, {Fox}, {Kamble}, {Kilic}, {Laskar}, {Morrell},
  {Nicholl}, {Paterson}, {Peters}, {Rojas-Bravo}, {Rest}, {Saario}, {Scolnic},
  {Soderberg}, {Strader}, {Villar}, \& {Williams}}]{Margutti2019}
{Margutti}, R., {Metzger}, B.~D., {Chornock}, R., {et~al.} 2019, ApJ, 872, 18,
  \dodoi{10.3847/1538-4357/aafa01}

\bibitem[{{McKernan} {et~al.}(2020){McKernan}, {Ford}, \&
  {O'Shaughnessy}}]{McKernan:2020lgr}
{McKernan}, B., {Ford}, K.~E.~S., \& {O'Shaughnessy}, R. 2020, MNRAS, 498,
  4088, \dodoi{10.1093/mnras/staa2681}

\bibitem[{{Metzger} {et~al.}(2024){Metzger}, {Hui}, \&
  {Cantiello}}]{Metzger:2024ujc}
{Metzger}, B.~D., {Hui}, L., \& {Cantiello}, M. 2024, ApJL, 971, L34,
  \dodoi{10.3847/2041-8213/ad6990}

\bibitem[{{Metzger} {et~al.}(2010){Metzger}, {Martinez-Pinedo}, {Darbha},
  {Quataert}, {Arcones}, {Kasen}, {Thomas}, {Nugent}, {Panov}, \&
  {Zinner}}]{Metzger2010}
{Metzger}, B.~D., {Martinez-Pinedo}, G., {Darbha}, S., {et~al.} 2010, MNRAS,
  406, 2650, \dodoi{10.1111/j.1365-2966.2010.16864.x}

\bibitem[{{Mikkola} \& {Merritt}(2008)}]{2008AJ....135.2398M}
{Mikkola}, S., \& {Merritt}, D. 2008, AJ, 135, 2398,
  \dodoi{10.1088/0004-6256/135/6/2398}

\bibitem[{Mooley {et~al.}(2022)Mooley, Anderson, \& Lu}]{Mooley:2022uqa}
Mooley, K.~P., Anderson, J., \& Lu, W. 2022, Natur, 610, 273,
  \dodoi{10.1038/s41586-022-05145-7}

\bibitem[{Most {et~al.}(2020)Most, Papenfort, Weih, \& Rezzolla}]{Most+20}
Most, E.~R., Papenfort, L.~J., Weih, L.~R., \& Rezzolla, L. 2020, MNRAS, 499,
  L82, \dodoi{10.1093/mnrasl/slaa168}

\bibitem[{{Naoz}(2016)}]{Naoz:2016cjb}
{Naoz}, S. 2016, ARA\&A, 54, 441, \dodoi{10.1146/annurev-astro-081915-023315}

\bibitem[{Narayan {et~al.}(2022)Narayan, Chael, Chatterjee, Ricarte, \&
  Curd}]{Narayan:2021qfw}
Narayan, R., Chael, A., Chatterjee, K., Ricarte, A., \& Curd, B. 2022, MNRAS,
  511, 3795, \dodoi{10.1093/mnras/stac285}

\bibitem[{Nathanail {et~al.}(2021)Nathanail, Most, \& Rezzolla}]{Nathanail+21}
Nathanail, A., Most, E.~R., \& Rezzolla, L. 2021, ApJL, 908, L28,
  \dodoi{10.3847/2041-8213/abdfc6}

\bibitem[{Palmese {et~al.}(2024)Palmese, Kaur, Hajela, Margutti, McDowell, \&
  MacFadyen}]{Palmese:2023beh}
Palmese, A., Kaur, R., Hajela, A., {et~al.} 2024, PhRvD, 109, 063508,
  \dodoi{10.1103/PhysRevD.109.063508}

\bibitem[{{Perley} {et~al.}(2019){Perley}, {Mazzali}, {Yan}, {Cenko}, {Gezari},
  {Taggart}, {Blagorodnova}, {Fremling}, {Mockler}, {Singh}, {Tominaga},
  {Tanaka}, {Watson}, {Ahumada}, {Anupama}, {Ashall}, {Becerra}, {Bersier},
  {Bhalerao}, {Bloom}, {Butler}, {Copperwheat}, {Coughlin}, {De}, {Drake},
  {Duev}, {Frederiks}, {Golkhou}, {Goobar}, {Heida}, {Ho}, {Horst}, {Hung},
  {Itoh}, {Jha}, {Kawai}, {Kuin}, {Kulkarni}, {Kumar}, {Kutyrev}, {Lee},
  {Littlejohns}, {Lunnan}, {Masci}, {Miller}, {Mogotsi}, {Murata}, {Neill},
  {Ngeow}, {Nugent}, {Ofek}, {Petrushevska}, {Rich}, {Ross}, {Roy}, {Rusu},
  {Shupe}, {Sollerman}, {Tachibana}, {Taddia}, {Tartaglia}, \&
  {Yu}}]{Perley2019}
{Perley}, D.~A., {Mazzali}, P.~A., {Yan}, L., {et~al.} 2019, MNRAS, 484, 1031,
  \dodoi{10.1093/mnras/sty3420}

\bibitem[{Pessi {et~al.}(2025)Pessi, Desai, Prieto, Kochanek, Shappee,
  Anderson, Beacom, Dong, Stanek, \& Thompson}]{Pessi:2025wht}
Pessi, T., Desai, D.~D., Prieto, J.~L., {et~al.} 2025, A\&A, 703, A34,
  \dodoi{10.1051/0004-6361/202556799}

\bibitem[{{Peters}(1964)}]{Peters:1964zz}
{Peters}, P.~C. 1964, PhRv, 136, B1224, \dodoi{10.1103/PhysRev.136.B1224}

\bibitem[{{Piro} \& {Pfahl}(2007)}]{Piro&Pfahl07}
{Piro}, A.~L., \& {Pfahl}, E. 2007, \apj, 658, 1173, \dodoi{10.1086/511672}

\bibitem[{{Rasio} \& {Ford}(1996)}]{1996Sci...274..954R}
{Rasio}, F.~A., \& {Ford}, E.~B. 1996, Sci, 274, 954,
  \dodoi{10.1126/science.274.5289.954}

\bibitem[{{Riess} {et~al.}(2022)}]{Riess:2021jrx}
{Riess}, A.~G., {et~al.} 2022, ApJL, 934, L7, \dodoi{10.3847/2041-8213/ac5c5b}

\bibitem[{{Schutz}(1986)}]{Schutz:1986gp}
{Schutz}, B.~F. 1986, Natur, 323, 310, \dodoi{10.1038/323310a0}

\bibitem[{Shibata {et~al.}(2024)Shibata, Fujibayashi, Lam, Ioka, \&
  Sekiguchi}]{Shibata:2023tho}
Shibata, M., Fujibayashi, S., Lam, A. T.-L., Ioka, K., \& Sekiguchi, Y. 2024,
  PhRvD, 109, 043051, \dodoi{10.1103/PhysRevD.109.043051}

\bibitem[{{Singer} {et~al.}(2016){Singer}, {Chen}, {Holz}, {Farr}, {Price},
  {Raymond}, {Cenko}, {Gehrels}, {Cannizzo}, {Kasliwal}, {Nissanke},
  {Coughlin}, {Farr}, {Urban}, {Vitale}, {Veitch}, {Graff}, {Berry},
  {Mohapatra}, \& {Mandel}}]{Singer:2016erz}
{Singer}, L.~P., {Chen}, H.-Y., {Holz}, D.~E., {et~al.} 2016, ApJL, 829, L15,
  \dodoi{10.3847/2041-8205/829/1/L15}

\bibitem[{{Smith} \& {Lissauer}(2009)}]{2009Icar..201..381S}
{Smith}, A.~W., \& {Lissauer}, J.~J. 2009, Icar, 201, 381,
  \dodoi{10.1016/j.icarus.2008.12.027}

\bibitem[{{Taddia} {et~al.}(2018)}]{Taddia:2017rvw}
{Taddia}, F., {et~al.} 2018, A\&A, 609, A136,
  \dodoi{10.1051/0004-6361/201730844}

\bibitem[{{Tagawa} {et~al.}(2021){Tagawa}, {Kocsis}, {Haiman}, {Bartos},
  {Omukai}, \& {Samsing}}]{Tagawa:2020qll}
{Tagawa}, H., {Kocsis}, B., {Haiman}, Z., {et~al.} 2021, ApJ, 908, 194,
  \dodoi{10.3847/1538-4357/abd555}

\bibitem[{{Tremaine}(2015)}]{2015ApJ...807..157T}
{Tremaine}, S. 2015, ApJ, 807, 157, \dodoi{10.1088/0004-637X/807/2/157}

\bibitem[{Tsokaros {et~al.}(2020)Tsokaros, Ruiz, \& Shapiro}]{Tsokaros+20}
Tsokaros, A., Ruiz, M., \& Shapiro, S.~L. 2020, ApJ, 905, 48,
  \dodoi{10.3847/1538-4357/abc421}

\bibitem[{{Vieira} {et~al.}(2020)}]{Vieira:2020lze}
{Vieira}, N., {et~al.} 2020, ApJ, 895, 96, \dodoi{10.3847/1538-4357/ab917d}

\bibitem[{{Villar} {et~al.}(2017){Villar}, {Guillochon}, {Berger}, {Metzger},
  {Cowperthwaite}, {Nicholl}, {Alexander}, {Blanchard}, {Chornock},
  {Eftekhari}, {Fong}, {Margutti}, \& {Williams}}]{Villar2017}
{Villar}, V.~A., {Guillochon}, J., {Berger}, E., {et~al.} 2017, ApJL, 851, L21,
  \dodoi{10.3847/2041-8213/aa9c84}

\bibitem[{{Watson} {et~al.}(2020)}]{Watson:2020iqj}
{Watson}, A.~M., {et~al.} 2020, MNRAS, 492, 5916, \dodoi{10.1093/mnras/staa161}

\bibitem[{{Woosley}(1993)}]{Woosley:1993wj}
{Woosley}, S.~E. 1993, ApJ, 405, 273, \dodoi{10.1086/172359}

\bibitem[{{Woosley} \& {Bloom}(2006)}]{Woosley:2006fn}
{Woosley}, S.~E., \& {Bloom}, J.~S. 2006, ARA\&A, 44, 507,
  \dodoi{10.1146/annurev.astro.43.072103.150558}

\bibitem[{Wu {et~al.}(2026)Wu, Most, Vu, Deppe, Kidder, Nelli, \&
  Throwe}]{Wu:2026hth}
Wu, J., Most, E.~R., Vu, N.~L., {et~al.} 2026, ApJL, 1004, L19,
  \dodoi{10.3847/2041-8213/ae743a}

\bibitem[{Yang {et~al.}(2019)Yang, Bartos, Haiman, Kocsis, Marka, Stone, \&
  Marka}]{Yang:2019okq}
Yang, Y., Bartos, I., Haiman, Z., {et~al.} 2019, ApJ, 876, 122,
  \dodoi{10.3847/1538-4357/ab16e3}

\bibitem[{{Ye} {et~al.}(2020){Ye}, {Fong}, {Kremer}, {Rodriguez}, {Chatterjee},
  {Fragione}, \& {Rasio}}]{Ye:2019xvf}
{Ye}, C.~S., {Fong}, W.-f., {Kremer}, K., {et~al.} 2020, ApJL, 888, L10,
  \dodoi{10.3847/2041-8213/ab5dc5}

\bibitem[{{Zevin} {et~al.}(2020){Zevin}, {Spera}, {Berry}, \&
  {Kalogera}}]{Zevin:2020gma}
{Zevin}, M., {Spera}, M., {Berry}, C. P.~L., \& {Kalogera}, V. 2020, ApJL, 899,
  L1, \dodoi{10.3847/2041-8213/aba74e}

\end{thebibliography}

\end{document}